\documentclass[12pt]{article}

\usepackage{newtxtext,newtxmath}
\usepackage{changes}

\usepackage{graphicx}

\usepackage{booktabs}

\usepackage[letterpaper,margin=1in]{geometry}

\usepackage[scale=0.94]{tgbonum}
\usepackage[mode=text]{siunitx}

\makeatletter
\DeclareTextCompositeCommand{\r}{OT1}{A}{%
  \leavevmode\vbox{%
    \offinterlineskip
    \ialign{\hfil##\hfil\cr\char23\cr\noalign{\kern-1.15ex}A\cr}%
  }%
}

\renewenvironment{abstract}
	{\quotation}
	{\endquotation}

\date{}

\newcommand{\beq}{\begin{equation}}
\newcommand{\eeq}{\end{equation}}

\makeatletter
\renewcommand{\fnum@figure}{\textbf{Figure \thefigure}}
\renewcommand{\fnum@table}{\textbf{Table \thetable}}
\makeatother

\usepackage{scicite}

\usepackage{url}

\def\scititle{
	Altermagnetism revealed by \\polarized neutrons in MnF$_2$
}
\title{\bfseries \boldmath \scititle}

\author{
        Quentin Faure$^{1\dagger}$,
        Dalila Bounoua$^{1\dagger}$,
        Victor Bal\'{e}dent$^{2,3\dagger}$,
        Arsen Gukasov$^{1}$,\and
        V. Ovidiu Garlea$^{4}$,
        Afonso Ribeiro$^{1}$,
        Jeffrey G. Rau$^{5}$,
        Sylvain Petit$^{1}$,
        Paul McClarty$^{1}$.\and
	\small$^{1}$Laboratoire L\'{e}on Brillouin, CEA, CNRS, Universit\'{e} Paris-Saclay, CEA Saclay, 91191 Gif-sur-Yvette, France.\and
	\small$^{2}$Universit\'{e} Paris-Saclay, CNRS, Laboratoire de Physique des Solides, 91405, Orsay, France.\and
	\small$^{3}$Institut universitaire de France (IUF)\and
	\small$^{4}$Neutron Scattering Division, Oak Ridge National Laboratory, Oak Ridge, Tennessee 37831, USA.\and
    \small$^{5}$Department of Physics, University of Windsor, 401 Sunset Avenue, Windsor, Ontario, N9B 3P4, Canada\and
	\small$^\dagger$These authors contributed equally to this work.
}

\begin{document} 

\maketitle

\begin{abstract} \bfseries \boldmath
Motivated by possible spintronics applications in antiferromagnets, it was recently observed that symmetry admits magnets that combine attractive features of both ferromagnets and antiferromagnets. These systems, dubbed {\it altermagnets}, have been the subject of intense study with direct spectroscopic evidence, from ARPES and RIXS, techniques reported in a handful of materials in the last year. Inelastic neutron scattering (INS) is one of the most powerful direct probes of magnetism and has recently been used to witness a splitting of magnon bands in MnTe that is compatible with altermagnetism although the nature and origin of the splitting remain to be fully characterized. However, the full power of neutron scattering for such systems comes from using polarized neutrons to measure the chirality of the magnon bands. Such a measurement provides a direct characterization of altermagnetism directly from the spin wave excitations. In this article, we present results on MnF$_2$ once thought to be an archetypal antiferromagnet. We present a polarized INS data that demonstrate the material is, in fact, altermagnetic. It had long been realized that the magnon bands in this material should have a weak splitting coming from the long-range dipolar coupling. Our data is the first to visualize this splitting directly. While the dipolar splitting on its own is not altermagnetic, using a domain biased sample, the data reveals a nonzero chirality in the neutron scattering cross section that reverses sign between the two magnon modes. It is this feature that clearly demonstrates altermagnetism in MnF$_2$. This finding highlights the potential for polarized INS to characterize altermagnets not least through its
exquisite sensitivity to fine-structure in the magnon spectrum.
\end{abstract}

\noindent

As condensed matter physics progresses, it has become possible to discern ever more subtle ways in which matter can organize itself. Whereas records of the magnetic properties of magnetite, an uncompensated magnet, appeared thousands of years ago, antiferromagnets are, to an untrained eye, magnetically inert. Their existence was irrefutably demonstrated only after the invention of neutron scattering techniques in the 1950's. Indeed, many states of matter that are the staple of modern quantum condensed matter research such as superconductivity, topological matter and fractionalization were beyond the reach of experimental science only a few decades ago. Altermagnetism, too, is subtle \cite{Smejkal2019,hayami2019,ssj2022,Smejkal2022b,jungwirth2025}. How else could it have escaped the attention of physicists until just a few years ago? Altermagnets have an underlying antiferromagnetic order parameter $-$ and all of the materials currently studied in this context had long been identified as mere antiferromagnets. But their distinctive feature, that sets them apart from their more conventional cousins, is a set of so-called {\it spin symmetries} \cite{ssj2022,Smejkal2022b} that are tied to the antiferromagnetism,  particular local multipolar order parameters \cite{bhowal2024,mcclarty2024} and, crucially, to a characteristic spin splitting of the band structure \cite{ssj2022,Smejkal2022b}. The first spectroscopic evidence for altermagnetism appeared only last year \cite{lee2024,liu2024,krempaski2024,osumi2024,hajlaoui2024,chilcote2024,amin2024,reimers2024,ding2024,zeng2024,li2025crsb,lu2024crsb} and considerable effort is being invested in finding new altermagnets and better ways to identify them. These efforts are being driven both to advance our fundamental understanding of magnetism and because altermagnets are promising candidate systems for future spintronics applications.

To date, the most well-established altermagnetic materials are MnTe \cite{lee2024,liu2024,krempaski2024,osumi2024,hajlaoui2024,chilcote2024,amin2024} and CrSb \cite{reimers2024,ding2024,yang2025,zeng2024,li2025crsb,lu2024crsb,biniskos2025systematic}. In both, there is evidence of an altermagnetic spin splitting of the band structures from angle resolved photo-emission spectroscopy data including direct evidence in MnTe coming from a spin-resolved study \cite{krempaski2024}. In other words, the altermagnetism has been visualized through the electronic degrees of freedom. However, altermagnetism is tied to the underlying magnetocrystalline symmetries and, as such, should appear in all degrees of freedom coupled to the magnetism. This includes the magnetic structure itself and the spin wave excitations.  Altermagnetism is most cleanly defined in the limit of zero spin-orbit coupling where the collinear magnetism in invariant under spin symmetries \cite{spinGroupsLO,spinPointLitvin,Liu,corticelli2022,schiff2023,xiao2023spin,ren2023enumeration,jiang2023enumeration} including a residual $U(1)$ spin rotation symmetry. In this limit, the magnons exhibit a pattern of chirality splitting analogous to the spin splitting of the electronic band structure first noted in Ref.~\cite{Smejkal2023} and explored further in \cite{Gohlke2023, cichutek2025,eto2025}. Although departures from the ideal limit break strict spin or chirality conservation, the energy splittings associated with altermagnetism imply that the phenomenon has a robustness that extends to real materials. Thus, a splitting of the magnon bands consistent with altermagnetism has been observed in MnTe using inelastic neutron scattering \cite{liu2024}. However, in order to make a definitive identification of altermagnetism from the magnons it is important to visualize the remnant chirality itself. Just such a measurement has been carried out using resonant inelastic X-ray scattering through the circular dichroism in CrSb \cite{biniskos2025systematic}. However, the energy resolution of RIXS is on the order of $10$ meV and insufficient to resolve any magnon splitting in this case.

In order to resolve both small splittings and the chirality, inelastic neutron scattering with polarization analysis is the method of choice \cite{mcclarty2025}. The method is perfectly tailored to probe magnetism on typical magnetic energy scales. As discussed in Ref.~\cite{mcclarty2025}, when unpolarized neutrons scatter from magnons in an altermagnet, they acquire a polarization along the scattering wavevector with a sign that directly reflects the chirality of the magnon mode. Then, through both the dispersion relations and polarization analysis, one may explore departures, originating from various possible magnetic anisotropies, from the limit where the magnon chirality is conserved.

In this paper, we report inelastic neutron scattering data on a candidate altermagnetic material MnF$_2$. The magneto-crystalline symmetries are compatible with altermagnetism in this material and the magnetic manganese has an essentially pure spin $5/2$ magnetic moment so spin-orbital effects are very weak, placing it among the more likely systems where magnons may carry a (weakly broken) chirality.  MnF$_2$ is well-known in the field of magnetism. Indeed, it was among the first materials in which an antiferromagnetic order parameter was identified in the early days of neutron scattering and it has been used in neutron training courses to teach students about canonical antiferromagnetism \cite{okazaki1964neutron,turberfield1965development,nikotin1969,holden1970,schweika2002,yamani2010,tseng2016,morano2024absence} $-$ a small irony that is indicative of the difficulty of detecting altermagnetism in this material. 

\begin{figure}[!]
\centering{\includegraphics[width=\linewidth]{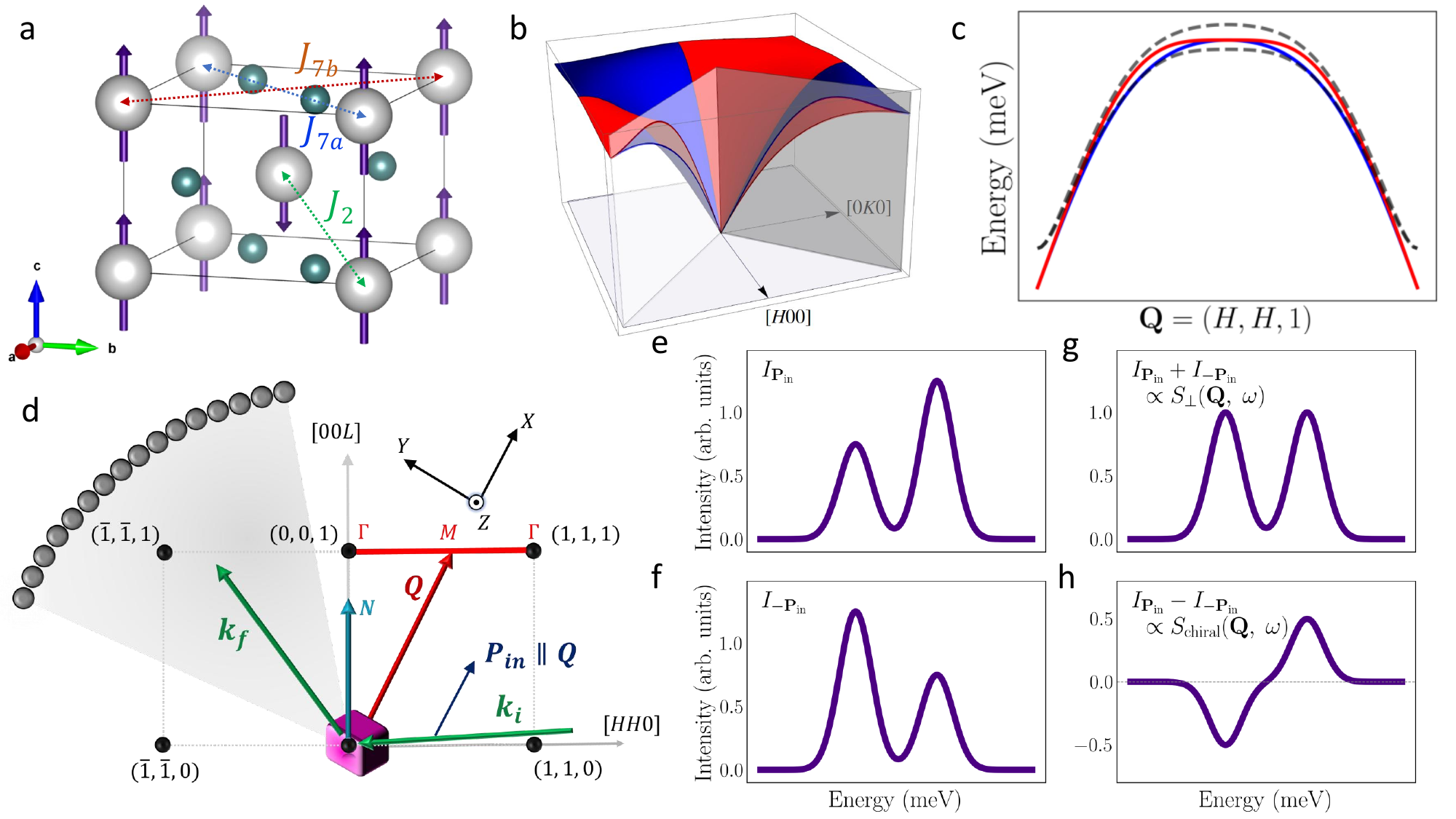}}
\caption{Plot depicting the origin and detection of altermagnetism in the spin waves of MnF$_2$. (a) The crystal and magnetic structure of MnF$_2$ with magnetic Mn$^{2+}$ ions on the sites of a body-centered tetragonal lattice and decorating fluoride ions that are crucial for the altermagnetism by breaking the symmetry connecting the magnetic sublattices down to a $C_{4}$ along the c axis, a $\boldsymbol{Q} = (1/2, 1/2, 1/2)$ translation and time reversal. This symmetry allows a magnon band splitting (b) in models with zero spin-orbit coupling. The pattern of magnon chiralities $\pm$ is shown in (b) through colors red ($+$) and blue ($-$) that is enforced by symmetry. Polarized neutron scattering is sensitive to this chirality. The maximal splitting is along $[HH0]$ in the middle of the zone (c, solid lines). Dipolar couplings, in the absence of an altermagnetic splitting, do not lead to a net chiral term but they do split the bands even at the zone boundary  (c, dashed lines). Panel (d) depicts the scattering geometry for the polarized inelastic neutron scattering experiment. The MnF$_2$ crystal is oriented in the $(HHL)$ plane, with its magnetic moments aligned along the Néel vector $\mathbf{N}$. An incident neutron beam, characterized by wavevector $\mathbf{k_{\rm i}}$ and polarization $\mathbf{P}_{\rm in}$, arrives at the sample under the condition $\mathbf{P}_{\rm in} \parallel \boldsymbol{Q} \parallel \boldsymbol{X}$, where $\boldsymbol{X}$ defines the $(\boldsymbol{X}, \boldsymbol{Y}, \boldsymbol{Z})$ Blume–Maleev coordinate system for polarized neutrons. The neutrons are then scattered into a final wavevector $\mathbf{k_{\rm f}}$ and detected by position-sensitive detectors. Identification of altermagnetism means detecting a non-vanishing chiral term in the neutron scattering that reverses sign between the two magnon branches. To investigate this: the scattering intensity is measured at constant momentum with ingoing polarization $\mathbf{P}_{\rm in}$ (e) and $-\mathbf{P}_{\rm in}$ (f). The usual unpolarized neutron cross section is the sum of these contributions (g) and the chiral contribution is isolated from their difference (h).}
\label{fig:intro}
\end{figure}

Previous inelastic neutron scattering data revealed a single magnon branch with a gap at the zone centre that was attributed to the long-range magnetostatic dipolar coupling \cite{okazaki1964neutron,turberfield1965development,nikotin1969,holden1970,schweika2002,yamani2010,tseng2016,morano2024absence}. While it was recognized that this coupling too should induce a small splitting between the two magnon modes \cite{okazaki1964neutron,turberfield1965development,nikotin1969}, this was below the resolution of the instruments in all previous experiments. In this work, our experimental data reveal the presence of a well-resolved magnon splitting with a maximum located at the zone boundary of about $0.2$~meV, compared to a $7$~meV total bandwidth. This originates primarily from the long-range dipolar coupling. 

The goal of the present study was to probe the magnon chirality. Whereas magnon chirality has previously been observed in uncompensated magnets \cite{nambu2020,jenni2022}, its detection in compensated magnets carries additional challenges. Indeed, for this purpose, it was necessary to have a sample that is magnetically ordered with a bias towards a single antiferromagnetic domain. From the diffraction data, we show that the two domains appear spontaneously upon cooling in a roughly 4:1 ratio, making the sample suitable for inelastic study with polarization analysis. A difference map of the intensities of opposed incident polarizations exhibits a clear sign reversal between the two magnon branches consistent with the presence of altermagnetism in this material. This study demonstrates that polarized neutron scattering can completely characterize altermagnetism directly from the magnetic degrees of freedom even in instances where the energy scale producing the altermagnetic splitting is significantly below that of perturbations that produce non-altermagnetic magnon band splittings. 

Before describing the details of the experiment on MnF$_2$ we briefly summarize the results expected from a polarized inelastic neutron scattering experiment on an {\it idealized} altermagnet $-$ that, as mentioned above, we take to mean an altermagnet with an exact $U(1)$ spin rotational symmetry in the magnetically ordered state. We further assume, for simplicity, that there are two magnetic sublattices. Then, the N\'{e}el vector $\mathbf{N}$ is the difference of the magnetizations on the two oppositely aligned sublattices. In order to be altermagnetic, $\mathbf{N}$ must transform as a non-trivial one dimensional irrep of the point group of the crystal. For the case of MnF$_2$, the two sublattices are related by a four-fold rotation about the tetragonal $c$ axis and a translation by half a unit in each direction. In the ideal altermagnet, the magnon bands are split over most of the Brillouin zone and the bands may be assigned chiralities $\pm 1$ that reverse under the same four-fold symmetry that connects the two sublattices. Now suppose we prepare the sample in a single magnetic domain and carry out an inelastic neutron scattering experiment in a half-polarized setting where we control the in-going neutron polarization $\mathbf{P}_{\rm in}$ and measure the overall scattered intensity. The result within linear spin wave theory is
\begin{align}
I(\mathbf{k},\omega) \propto & \sum_{n} \left[ \frac{1}{2} \left(  1 + \left( \hat{\mathbf{k}}\cdot\hat{\mathbf{N}} \right)^2 \right)  + (-)^n \left( \mathbf{P}_{\rm in}\cdot \hat{\mathbf{k}} \right) \left( \hat{\mathbf{k}}\cdot \hat{\mathbf{N}} \right) \right] C_{\mathbf{k}} \delta\left(\omega - \epsilon^{(n)}_{\mathbf{k}}\right).
\end{align}
with $\hat{\mathbf{k}}$ the wave vector and $C_{\rm{\mathbf{k}}}$ the one magnon band intensity~\cite{mcclarty2025}.
The first term in square brackets is the usual unpolarized neutron scattering intensity which is the same in the two split modes. The second (chiral) term in square brackets has a polarization dependence and a sign that depends on the mode index $n$. This term is present whenever the scattering wavevector has a component along the in-going neutron polarization and along the staggered magnetization. It is this term that provides a direct measurement of the altermagnetic features. In practice one may make a measurement of the intensity for $+\mathbf{P}_{\rm in}$ and for $-\mathbf{P}_{\rm in}$, subtract the two in order to isolate the chiral term in the cross section.  

Now we are in a position to describe the experimental details. The work was carried out on a single crystal of MnF$_2$ which has tetragonal space group P4$_2$/mnm (\# $136$) with pure spin $S=5/2$ manganese ions on $2a$ sites and nominally non-magnetic fluoride ions on $4f$ sites. The material magnetically orders below $T_N\approx 67$ K into a simple collinear antiferromagnetic structure with moments oriented parallel and antiparallel to the $c$ axis [Fig.~\ref{fig:intro}(a)]. As mentioned above, the altermagnetic features of MnF$_2$ originate from the four-fold rotation (combined with time-reversal and a half-translation) connecting the two magnetic sublattices. This directly leads to the prediction of d-wave altermagnetism with a pattern of chirality splitting indicated in Fig.~\ref{fig:intro}(c). We expect to witness the chirality through polarized inelastic neutron scattering provided the sample orders into a majority single domain configuration. 

In order to assess the magnetic domain population in the sample, we make use of the fact that the elastic neutron cross section at reflection $\mathbf{G}$ takes the form $I_{\mathbf{P}_{\rm in}}(\mathbf{G}) \propto \vert F_{\rm nuc}\vert^2 + \vert F_{\rm mag} \vert^2 + \left( \mathbf{P}_{\rm in}^\perp \cdot \hat{\mathbf{N}}\right) {\rm Re}\left[ F_{\rm nuc} F^*_{\rm mag} \right]$ where $F_{\rm nuc}$ and $F_{\rm mag}$ are, respectively, the nuclear and magnetic structure factors and $\mathbf{P}_{\rm in}^\perp$ is the component of the in-going neutron polarization perpendicular to the scattering wavevector \cite{supp}. The time-reversal odd nuclear-magnetic interference term has non-vanishing amplitude at $(H,K,L)$ reflections for $H+K+L$ odd. To isolate this term, the flipping ratio at the $(210)$ Bragg position, $I_{\mathbf{P}_{\rm in}}/I_{-\mathbf{P}_{\rm in}}$ was determined with $\mathbf{P}_{\rm in}$ parallel to the $c$ axis. The single crystal was found to have an overall mosaicity of $\approx 3^{\circ}$, arising from three distinct grains. From flipping ratio measurements, the domain population of each grain was determined \cite{supp} at $2$ K after cooling slowly through the magnetic transition. The result is that this protocol was sufficient on our samples to obtain a magnetic configuration with $85\pm 5 \%$ of the volume in a single domain. 

\begin{figure}[!]
\centering{\includegraphics[width=\linewidth]{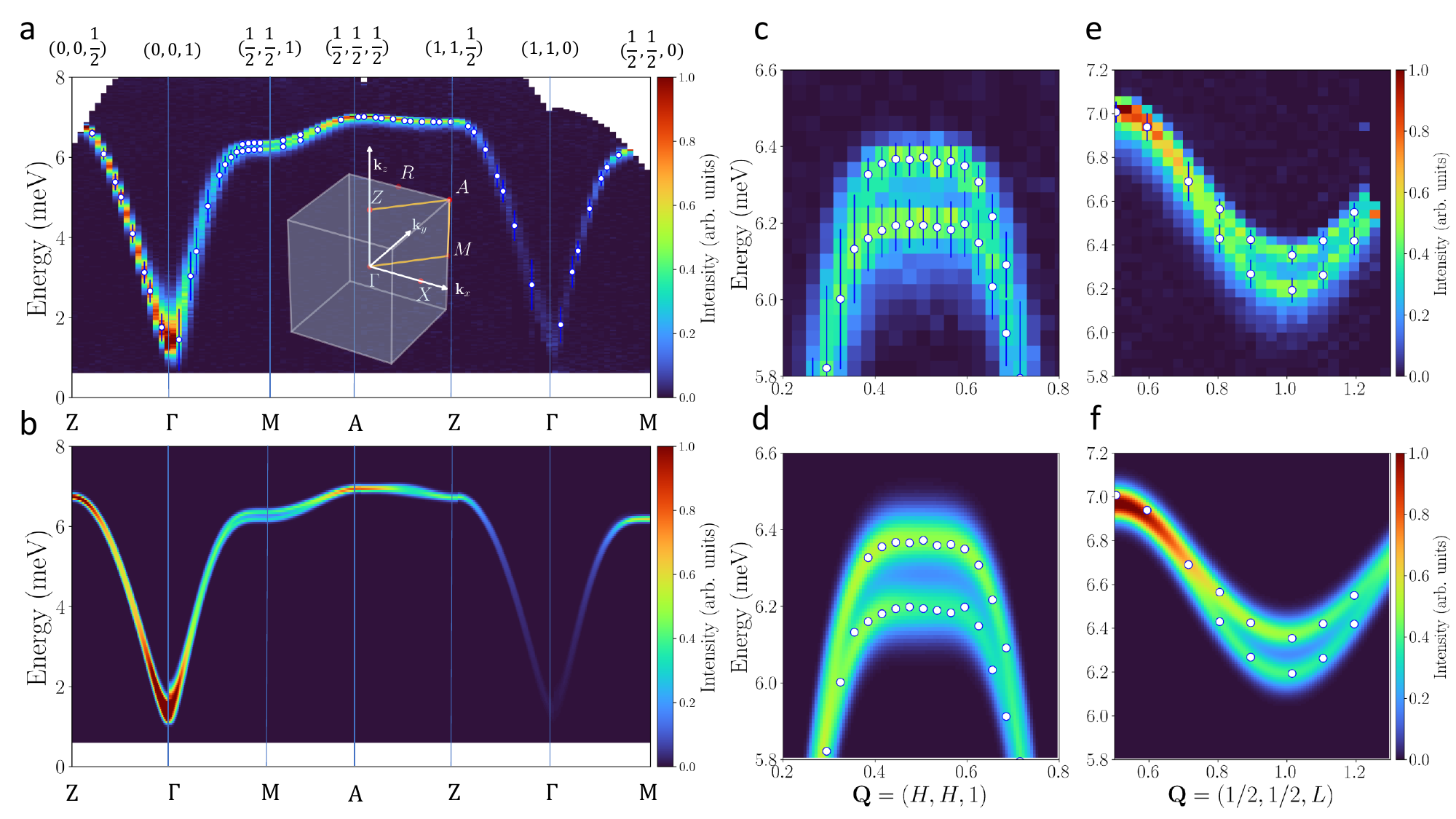}}
\caption{Unpolarized inelastic neutron spectra of MnF$_2$. (a) inelastic neutron spectra along high-symmetry directions within the Brillouin zone (inset) obtained with an incident energy of $E_i = 9$~meV. Blue circles shows energies of the magnetic modes obtained from fits of the data using a sum of Gaussians. (b) Calculated spin-wave spectra obtained from parameters (within the text) obtained from global fits of the data. (c,e) Zoom onto the spin wave spectrum of MnF$_2$ along $\boldsymbol{Q} = (H, H, 1)$ and $\boldsymbol{Q} = (1/2, 1/2, L)$ where the dipolar splitting is revealed. (d,f) Corresponding calculated spin-wave spectra. All data are integrated over $0.04$~meV steps along energy and $0.03$~r.l.u. steps in $\boldsymbol{Q}$ along the plotted axis. The transverse-$\boldsymbol{Q}$ averaging window is $0.1$ r.l.u. in all cases. Calculations were based on a Heisenberg exchange model with long-range dipolar couplings. The exchange couplings obtained by fitting the experimental data are:
$J_1 = -0.075(2)$ {\rm meV}, $J_2 = 0.287(3)$  {\rm meV}, $J_3 = -0.012(1)$  {\rm meV}, $J_4 = -0.001(1)$ {\rm meV}, $J_5 = 0.008(2)$  {\rm meV}, $J_6 = 0.001(2)$  {\rm meV}, $J_{7a} = -0.006(3)$ {\rm meV}, $J_{7b} = -0.002(3)$  {\rm meV}.
}
\label{fig:unpol}
\end{figure}

We now turn to the inelastic neutron scattering results obtained on the HYSPEC spectrometer at SNS (ORNL) with measurements made in the $(HHL)$ scattering plane. Panel (a) in Fig.~\ref{fig:unpol} shows unpolarized scattering intensity along various high symmetry line segments in momentum space (the paths indicated in the inset). The main feature visible in this plot is a dispersive branch of excitations with modulated intensity extending from about $1$ meV at the zone center up to about $7$ meV at the zone corners, consistent with previous measurements. The HYSPEC spectrometer at $9$ meV incident energy provides an energy resolution of about $100\mu$eV at an energy transfer of $6$meV. Close inspection of Fig.~\ref{fig:unpol}(a) shows that the main dispersive branch actually consists of two weakly split modes. The splitting is most clearly visible at $(1/2,1/2,1)$. Panels (c) and (e) show the experimental dispersion relations passing through this point, respectively along $(H,H,1)$ and $(1/2,1/2,L)$ showing clearly two modes with a maximum splitting of about $178(3)\mu$eV.  

As MnF$_2$ has two magnetic sublattices, two spin wave modes are expected. Moreover, the magneto-crystalline symmetries do not protect their degeneracy. However, as the manganese ions carry pure spin $5/2$ any exchange anisotropies should be very weak. Even so, the gap $\Delta$ at the zone center points to departures from the pure Heisenberg limit. Previous work recognized that this gap originates mainly from the long-range magnetostatic dipolar coupling \cite{okazaki1964neutron,turberfield1965development,nikotin1969}. The magnitude of the dipolar coupling, $D_{\rm nn}$, between nearest neighbor moments is fixed by the moment size and is of the order of a few tens of $\mu$eV. The magnitude of $\Delta$ is compatible with the small dipolar coupling as $\Delta \sim \sqrt{J D_{\rm nn}}$. However, the maximum splitting between the magnon branches induced by the dipolar coupling is roughly on the scale of $D_{\rm nn}$ which is therefore fine-structure compared to the magnon bandwidth. Indeed a calculation reveals that the dipolar splitting is maximum on the zone boundary with magnitude $178(3)$~$\mu$eV compatible with the HYSPEC data. 

\begin{figure}[!]
\includegraphics[width=\linewidth]{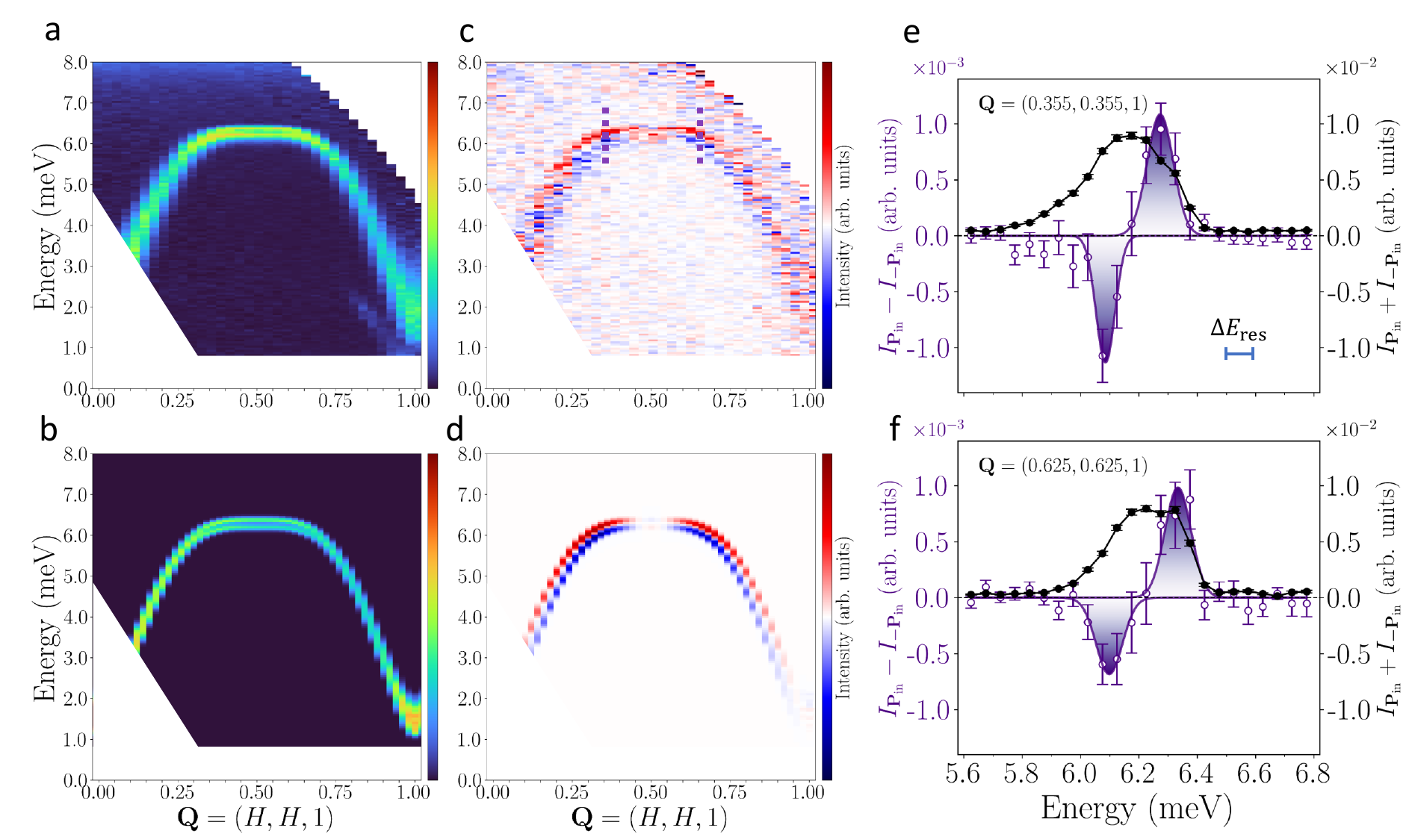}
\caption{Chirality of magnon bands in MnF$_2$ revealed by polarized neutrons. (a) Sum of spin wave spectra measured along $\boldsymbol{Q} = (H, H, 1)$ with polarized neutrons $+\mathbf{P}_{\rm{in}}$ and $-\mathbf{P}_{\rm{in}}$. The sum is only relative to the unpolarized cross section. (b) Corresponding calculated spin-wave spectra obtained from fitted parameter of unpolarized neutrons spectra. (c) The difference of spectra, relative only to the chiral part of the cross section. Dashed vertical purple lines represent constant-$\boldsymbol{Q}$ energy slices shown in (e-f). (d) Calculated chiralities of the magnon bands obtained from fitted parameters. The pronounced broadening of the magnon bands observed experimentally, compared to the calculations—particularly in regions of steep dispersion (0–0.3~r.l.u. and 0.7–1~r.l.u.) is likely due to a combination of data integration along perpendicular directions and the mosaic spread of the sample ($\simeq 3^{\circ}$). (e-f) Constant-$\boldsymbol{Q}$ energy scans obtained  from experimental sum (black points) and difference (purple points) at $\boldsymbol{Q} = (0.355, 0.355, 1)$ (e) and $\boldsymbol{Q} = (0.625, 0.625, 1)$ (f). Black curves are guide to the eye. Purple curve are results from a double Gaussian fit. All data herein are integrated over 0.05~meV steps along energy and 0.03~r.l.u. steps in $\boldsymbol{Q}$ along the plotted axis. The transverse-$\boldsymbol{Q}$ averaging window is 0.1 r.l.u. in both cases. The blue line in (e) denotes the energy resolution $\Delta E_{\rm{res}}$ which is calculated to be around 0.1~meV at 6~meV. This value approximately corresponds to the full width at half maximum (FWHM) of each fitted Gaussian; in the fits, the minimum allowed FWHM was constrained by $\Delta E_{\rm{res}}$.}
\label{fig:pol}
\end{figure}

We extracted dispersion points from the data by fitting cuts at constant energy (for steep dispersions) and constant momentum (otherwise). Fig.~\ref{fig:unpol}(a) shows a selection of these points overlaid onto the scattering intensity plots showing that the extracted points faithfully capture the measured intensities. We parameterized the data by carrying out a multiboson (or flavor wave) calculation based on a Hamiltonian $H=H_{\rm ex} + H_{\rm dip}$ where $H_{\rm ex}$ consists of Heisenberg exchange couplings $J_n$ between neighboring Mn ions at shell $n$ for $n=1,\ldots,7$ \footnote{This is equivalent to linear spin wave theory up to a scale depending on the spin length.}. The cutoff was fixed at $7$ because for shorter range Heisenberg couplings the symmetry of the model is higher than that of the lattice. Only when $J_{7}$ is included do the interactions capture the correct space group symmetry of MnF$_2$. This feature of the model is crucial as altermagnetic splittings in the zero spin-orbit coupled limit may arise {\it only} once the model has this symmetry. In fact, the symmetry lowering at $7$th neighbor is correlated to the fact that at this range there are two distinct couplings denoted $J_{7a}$ and $J_{7b}$ indicated in Fig.~\ref{fig:intro}(a). In order for altermagnetic splittings to appear, the four-fold symmetry of the body-centred tetragonal sublattice occupied by Mn ions must be broken by setting  $J_{7a} \neq J_{7b}$. In addition to the exchange, we include also the long-range dipolar interaction $H_{\rm dip}$ which has a fixed coupling strength. As we have no {\it a priori} knowledge of the exchanges we carry out a least squared fit to the data including $8$ free parameters (corresponding to the Heisenberg couplings just described) assuming spin-orbital anisotropies to be negligible. Further details of the data processing and fitting procedure may be found in the supplementary section~\cite{supp}. 

Fig.~\ref{fig:unpol}(b,d,f) show the results of calculations of the unpolarized neutron scattering intensity based on parameters extracted from the fitting procedure. The dispersion relations including the splitting of the modes are all very well captured by the fit and the calculation. The intensities are broadly in agreement between the experiment and the calculation $-$ for example the modulation along $(1/2,1/2,L)$ and the strong and weak intensities at the zone centers at $(0,0,1)$ and $(1,1,0)$ respectively. There is an apparent quantitative difference along the $\Gamma M$ path where the calculation has monotonically decreasing intensities away from the zone center whereas the experiment has intensities that increase again towards $(1/2,1/2,1)$. Additional comparisons between the experiment and calculations are given in the supplementary section \cite{supp}.

In order to parameterize the uncertainty in the exchange couplings, we carried out fits to the experimental dispersion points together with  Gaussian noise with variance on each point taken from the experimental peak widths. We focus here on features of the fits to noisy data that relate to altermagnetic couplings $J_{7a}$ and $J_{7b}$. We find that these couplings are constrained to be much smaller than the nearest neighbor exchange, $J_2$, between the two magnetic sublattices as might be expected especially in a magnetic insulator like MnF$_2$. In the fits shown in the figures we take $J_2=0.287$ meV while $J_{7a}=-0.006$ meV and $J_{7a}=-0.002$ meV. We find that $J_{7a}$ and $J_{7b}$ are strongly anti-correlated. 
Therefore the data are fully compatible with $J_{7a}-J_{7b}$ nonzero and also on the order of roughly  $10$~$\mu$eV~\cite{supp}.  

Having achieved some further understanding of the magnetic couplings in MnF$_2$ based on the HYSPEC data, we present the results of the polarized inelastic neutron scattering experiment whose goal was to find evidence of a nonzero chiral term in the cross section indicative of altermagnetism. The experiment was carried by fixing $\mathbf{P}_{\rm in}$ parallel to $\boldsymbol{Q}=(1/2,1/2,1)$ and measuring scattering intensity \cite{supp}. Then, a similar measurement was performed with fixed $-\mathbf{P}_{\rm in}$. The standard inelastic neutron scattering intensity is proportional to the sum of these measurements and the chiral term is isolated from the difference [Fig.~\ref{fig:intro}(e-h)]. The sum and difference maps along $(H,H,1)$ are shown in Figs.~\ref{fig:pol}(c) and (e) respectively with the former showing a pronounced splitting in the vicinity of the zone boundary. Focussing on panel (e), we observe a clear signal correlated to the magnetic excitations with a sign reversal between the upper and lower modes. Cuts at constant momentum \cite{supp} reveal that the magnitude of the chiral term has a maximum at about $H=0.35$ \cite{supp} at that it goes to zero, within errors, at the boundary $H=1/2$. Taken on its own with no further analysis, Figs.~\ref{fig:pol}(e) provides a compelling signature of altermagnetism in MnF$_2$. The magnetocrystalline symmetries of MnF$_2$ guarantee that the sign of the chiral term reverses when passing from $(H,H,1)$ to $(H,-H,1)$. 

It is interesting to examine possible microscopic origins of the chiral neutron scattering intensity in MnF$_2$. Let us first consider a model where the exchange constants have the constraint that $J_{7a}=J_{7b}$ so that $H_{\rm ex}$ alone leaves the magnon bands degenerate. Introduction of the long-range dipolar coupling lifts the degeneracy over most of the zone. The dipolar interaction in the model couples point-like magnetic moments living on a body-centered tetragonal lattice. This model has a fold-fold rotation symmetry leaving the magnetic sublattices invariant. This symmetry is incompatible with altermagnetism. Indeed, the dipolar coupling maximally mixes chiralities so that the chiral term in the neutron cross section vanishes. Starting from this model we may now tune $\delta J_7 \equiv \vert J_{7a}-J_{7b}\vert$ to be nonzero. For infinitesimal $\delta J_7$, a finite chiral term is present although the presence of the dipolar coupling means that the magnon chirality is not a good quantum number. This result shows concretely how altermagnetism can persist even into regimes where the altermagnetic splitting is small compared to those caused by chirality breaking anisotropies. 

We integrate the chiral contribution to the intensity finding that it is between $5\%$ and $11\%$ of the total intensity depending on the wavevector. We confirm that this ratio is compatible with the fact that the fits to the dispersion relations produce parameters with $\delta J_7 < 5\mu$eV. In other words, the polarization signal directly probes a significantly smaller energy scale than is directly measured through the dispersion relations. Therefore one message of this work, that is potentially important beyond the scope of altermagnetism, is that polarized neutron intensity may be sensitive to fine-structure in chiral magnon spectra to which more standard probes may be insensitive. This is analogous to magnetic circular dichroism, where a small Zeeman splittings due to an applied magnetic field can be resolved due to their derivative like spectrum~\cite{buckingham1966magnetic}.

As we have established, polarized neutron scattering provides a powerful means of detecting altermagnetism (i) by focussing directly on the magnetic degrees of freedom (ii) through its sensitivity to perturbations away from the idealized altermagnetic limit on the sub-meV scale through the dispersion relations (iii) through the fact that the cross section is known precisely in terms of underlying magnetic correlators, and, crucially,  (iv)  through its ability to measure the key feature of altermagnetism namely the chirality of the magnons. We have demonstrated these features through measurements that establish altermagnetism in MnF$_2$. These results provide a concrete benchmark for understanding altermagnetic spin excitations and will guide the exploration of altermagnetism and magnon chirality across a wide range of condensed matter systems.

\bibliography{refs_mnf2}
\bibliographystyle{sciencemag}


\section*{Acknowledgments}

\paragraph*{Funding:}
This research used resources at the Spallation Neutron Source, a DOE Office of Science User Facility operated by the Oak Ridge National Laboratory. The beam time was allocated to HYSPEC (BL-14B) on proposal number IPTS-34497. We thank G\'erard Lapertot for  having grown the MnF$_2$ single crystals, and Fr\'{e}deric Bourdarot and Oscar Fabelo for providing the samples. PM acknowledges funding from the CNRS. AR acknowledges a PhD studentship from the CEA. VB, DB, QF, AG, SP and PM acknowledge support from the French Federation of Neutron Scattering (2FDN). JR acknowledges funding from the Natural Sciences and Engineering Research Council of Canada (NSERC). VB acknowledges the MORPHEUS platform at the Laboratoire de Physique des Solides for sample selection, characterization and alignment for the neutron experiment. 

\paragraph*{Author contributions:}
QF, DB, VB, AG, PM, JR designed the neutron scattering experiment. VB aligned and prepared the samples. DB, QF, AG, VOG carried out the experiment. AR and PM carried out the theoretical part with contributions to the spin wave calculations from QF and SP. All authors contributed to the analysis of the data and the writing of the manuscript.

\paragraph*{Competing interests:}
There are no competing interests to declare.

\paragraph*{Data and materials availability:}
Data are available from the authors upon reasonable request.


\subsection*{Supplementary materials}
Materials and Methods\\
Supplementary Text\\
Figs. S1 to S10\\
Tables S1 to S3\\
References \textit{(47-\arabic{enumiv})}\\ 


\newpage


\renewcommand{\thefigure}{S\arabic{figure}}
\renewcommand{\thetable}{S\arabic{table}}
\renewcommand{\theequation}{S\arabic{equation}}
\renewcommand{\thepage}{S\arabic{page}}
\setcounter{figure}{0}
\setcounter{table}{0}
\setcounter{equation}{0}
\setcounter{page}{1} 


\begin{center}
\section*{Supplementary Materials for\\ \scititle}

	Quentin Faure,
	Dalila Bounoua,
        Victor Bal\'{e}dent,
	Arsen Gukasov,
        V. Ovidiu Garlea,
        Afonso Ribeiro,
        Jeffrey G. Rau,
        Sylvain Petit,
        Paul McClarty

\end{center}

\subsubsection*{This PDF file includes material on:}
Sample characterization \\
Experimental techniques\\
Domain population measurements \\
Unpolarized inelastic scattering measurements \\
Polarized inelastic scattering measurements \\
Minimal model \\
Spin wave theory and neutron cross section \\
Polarized neutron cross section \\
Parametrization of the dispersion relations \\
{\it A priori} constraints on the anisotropic couplings \\
Figures S1 to S8\\
Tables S1 to S3

\newpage


\subsection*{Materials and Methods}

\subsubsection*{Sample characterization}

The inelastic neutron scattering experiment was carried out on a MnF$_2$ single crystal synthesized using the Bridgman technique with a size of approximately $5 \times 5 \times 7$ mm. MnF$_2$ crystallizes in the P4$_2$/mnm space group (space group $\# 136$) with lattice parameters of $a = b = 4.87$~\r{A}\ and $c = 3.31$~\AA\ at 2~K. It undergoes a magnetic transition at $T_N \simeq 67$~K to a long range ordered antiferromagnetic state where  Mn$^{2+}$ ($S = 5/2, L = 0$) spins align antiferromagnetically along the c-axis with a propagation vector $\mathbf{k} = (0, 0, 0)$. 
This single crystal has been found to contain three different grains with a mosaic spread of $\approx3^\circ$ (see Fig.~\ref{fig:Domains}). 

\subsubsection*{Experimental Techniques}

Neutron scattering measurements were performed using the HYSPEC time-of-flight spectrometer at the Spallation Neutron Source, Oak Ridge National Laboratory (USA). The instrument was operated in three primary configurations:

\begin{enumerate}
    \item \textbf{Polarized neutron diffraction mode} using a polarized incident beam without polarization analysis, was employed to determine the antiferromagnetic domain population in MnF$_2$. The neutron polarization was oriented along the z-axis, perpendicular to the scattering plane, using a 3D Helmholtz-like coil to control the magnetic field at the sample position.
    
    \item \textbf{Unpolarized inelastic neutron scattering mode} using a polarized incident beam without polarization analysis of the scattered beam, was used to measure spin wave spectra. 
    
    \item \textbf{Polarized inelastic neutron scattering} was utilized to detect chiral correlations associated with altermagnetism.  In this setup, the neutron polarization was oriented within the scattering plane, parallel to the scattering vector at the energy transfer of interest.
\end{enumerate}

\subsubsection*{Domain population measurements}

To evaluate the domain population the sample was oriented with the $(HK0)$ horizontal plane and mounted on an aluminum holder, placed inside an orange cryostat, and cooled to $2$~K. The Bragg diffraction pattern was recorded using a detector array that covers an angular range of $60^{\circ}$ in the horizontal scattering plane and $\pm7^{\circ}$ vertically. The entire bank can be rotated about the sample to provide a $2\theta$ coverage from $2^{\circ}$ to $135^{\circ}$. 

\begin{figure}[t]
    \centering
    \includegraphics[width=.8\columnwidth]{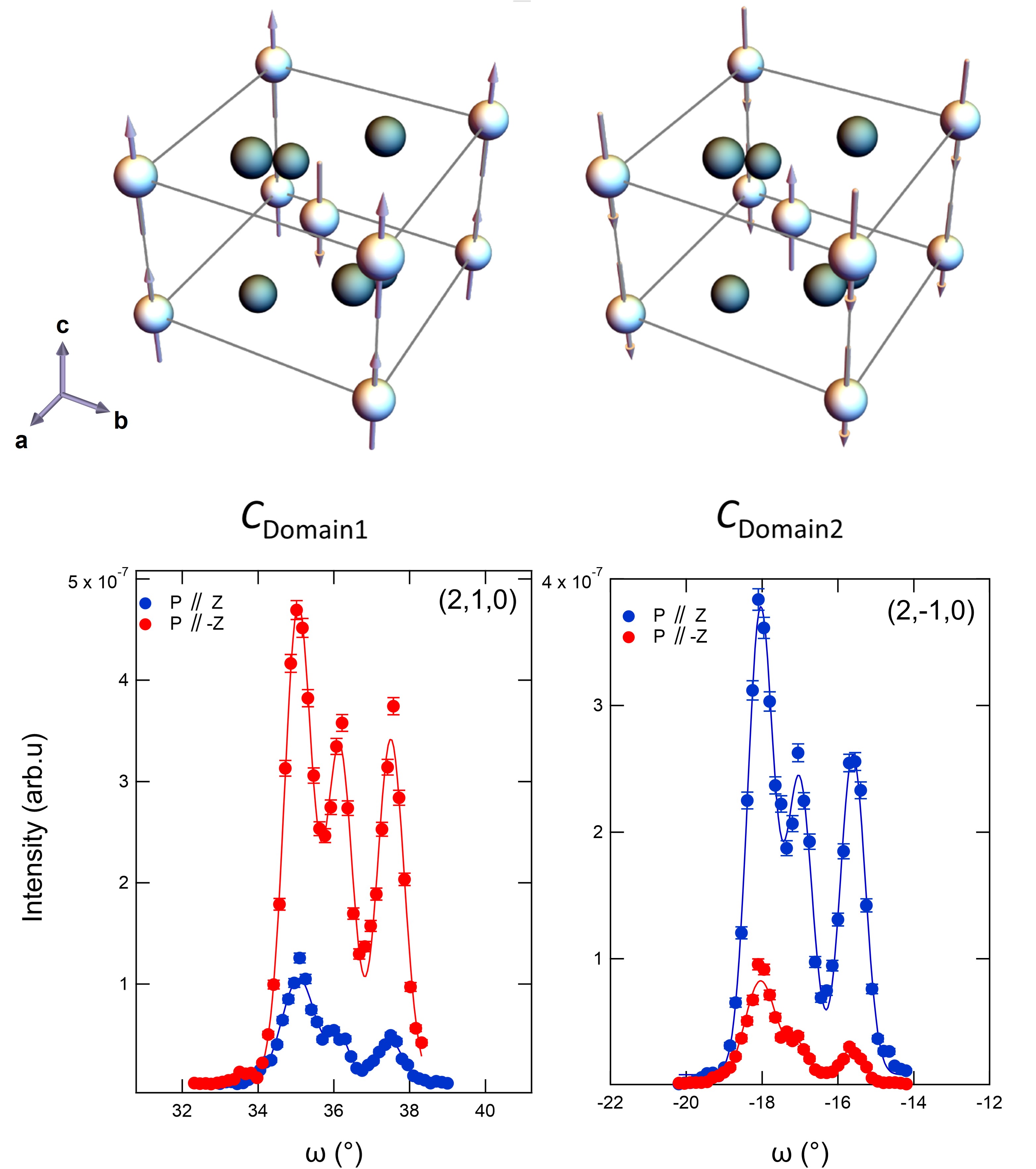}
    \caption{Top: The two magnetic domains $C_{\rm Domain 1}$ and $C_{\rm Domain 2}$ in MnF$_2$. Bottom: Rocking scans across the $(2,1,0)$ and $(2,-1,0)$ reflections with the neutron spin polarization $\mathbf{P}_{\rm in} \parallel \boldsymbol{Z}$ and $\mathbf{P}_{\rm in} \parallel - \boldsymbol{Z}$}
    \label{fig:Domains}
\end{figure}

The first part of the experiment aimed to determine the magnetic domain population in the MnF$_2$ single crystal. It was found that, at 2~K, approximately $85 \pm 5$\% of the crystal volume consisted of a single magnetic domain. Although the mosaic spread of the sample is large $\approx3^\circ$, the flipping ratios determined on the three grains shown in Fig.~\ref{fig:Domains} are found to contribute similarly to the overall domain population. This was determined using the flipping ratio method, as described by Nathans et al.~\cite{nathans1963}.

The method involves Bragg diffraction of a monochromatic, polarized neutron beam on a single crystal placed in a magnetic field. The experiment measures the intensity of Bragg peaks for neutron spins aligned parallel ($I^+$) and antiparallel ($I^-$) to the magnetic field. The key experimental observable is the flipping ratio, defined as:
\[
R = \frac{I^+}{I^-}.
\]
For centrosymmetric crystals, and when the magnetic moments are aligned parallel to the polarization direction, the intensities can be written as:
\[
I^\pm = F_N^2 \pm 2 P_i F_N F_M + F_M^2
\]
where $F_N$ and $F_M$ are the nuclear and magnetic structure factors, respectively, and $P_i$ is the incident neutron polarization.

In the case of MnF$_2$, the absolute values of the nuclear and magnetic structure factors for the (210) reflection are nearly equal at low temperatures. Under this condition, the flipping ratio simplifies to:
\[
R = \frac{1 + P_i (2\alpha - 1)}{1 - P_i \epsilon (2\alpha - 1)}.
\]
Here, $\alpha$ represents the fraction of the crystal in one of the two domain states, and $\epsilon$ is the efficiency of the spin flipper in reversing the neutron polarization. Both $P_i$ and $\epsilon$ were calibrated using polarization analysis with a supermirror analyzer and a standard Heusler crystal of known polarization efficiency. Additionally, the polarization efficiency was verified using a pure nuclear reflection from the sample.

For these polarized diffraction measurements, the crystal was mounted with the crystallographic $c$-axis and the incident neutron polarization $P_i^Z$ aligned along the vertical ($Z$) direction.

It is important to note that, unlike conventional flipping ratio measurements in paramagnetic materials where the magnitude of the magnetic field plays a crucial role in altermagnets, only the orientation of the neutron spin with respect to the magnetic moments is relevant. Consequently, a standard XYZ coil system on HYSPEC was employed to generate a small, vertically oriented guide field of approximately 20 Gauss to maintain the desired polarization direction.

Flipping ratios were then measured for the (2,1,0) and (2,\(-1\),0) reflections. These measurements enabled the determination of the domain population balance (see Table).
  
\begin{table}[htbp]
\centering
\caption{Measured flipping ratios and calculated domain fractions for selected Bragg reflections.}
\begin{tabular}{llcc c}
\toprule
  &Reflection& Grain 1& Grain 2& Grain 3\\ 
\midrule
 & $R_{(2,1,0)}$  & $4.54 \pm 0.20$ & $7.89 \pm 0.65$ & $7.75 \pm 0.54$ \\
& $C_{\rm Domain 1}$ & $\approx 18$ \%& $\approx 11$ \% & $\approx 11$ \% \\
\midrule
 & $R_{(2,-1,0)}$  & $0.22 \pm 0.01$ & $0.12 \pm 0.01$ & $0.1 \pm 0.01$ \\
 & $C_{\rm Domain 2}$  & $\approx 82$ \%& $\approx 89$ \%& $\approx 90$ \%\\\bottomrule
\end{tabular}
\end{table} 
 As expected, the table also confirms that reflections located in successive quadrants are related by the symmetry relation
\[
R_{hkl} = \frac{1}{R_{h\bar{k}l}}.
\]
We emphasize that this measurement not only quantifies the extent to which the sample exhibits a single antiferromagnetic domain, but also determines the sign of $\alpha$. This, in turn, specifies the direction of the spins on the magnetic sublattice centred on, $\mathbf{M}_1(0,0,0)$, (i.e. those with fluoride ions at $\pm(x,x,0)$) for the (210)-type reflection. Consequently, the sign of the Néel vector
\[
\mathbf{N} = \mathbf{M}_1(0,0,0) = -\mathbf{M}_2\left(\frac{1}{2}, \frac{1}{2}, \frac{1}{2}\right)
\]
is also established.

\subsubsection*{Unpolarized Inelastic Scattering Measurements}

\begin{figure}[t]
    \centering
    \includegraphics[width=0.6\columnwidth]{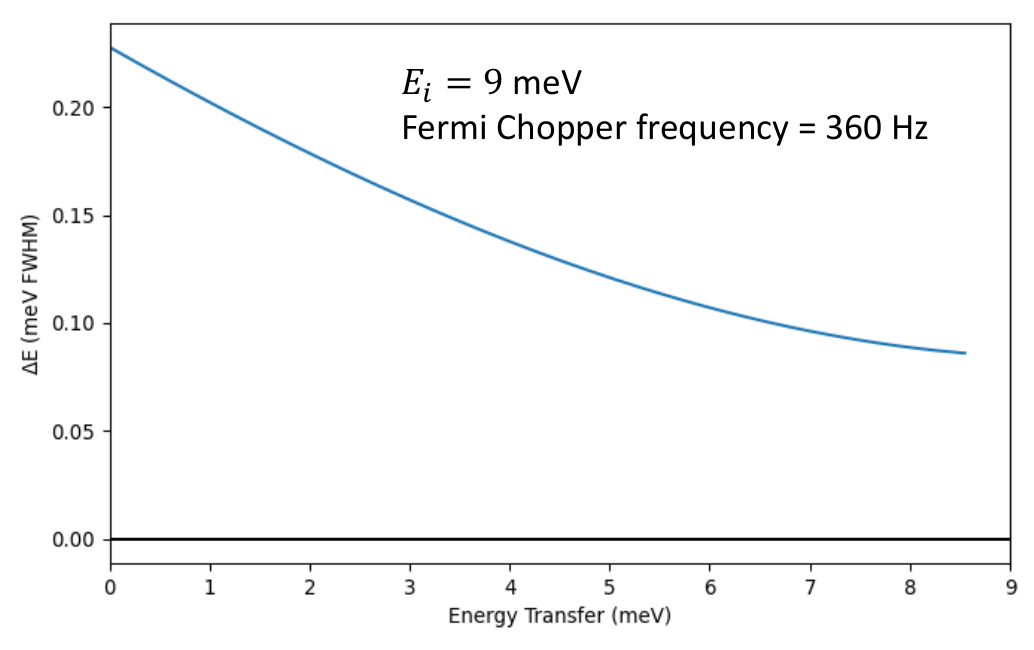}
    \caption{Calculated energy resolution for an incident energy $E_i = 9$~meV and a Fermi chopper frequency of 360~Hz.}
    \label{fig:Eresolution}
\end{figure}

Wide-angle time-of-flight (TOF) spectrometer HYSPEC features a detector bank that registers neutrons with scattered wave vectors $\mathbf{k}_f$ filling a large volume of phase space, spanning a broad range of directions and magnitudes. The sample was mounted with the $[1\bar{1}0]$ axis vertically oriented, providing access to the $(HHL)$ scattering plane.
Subsequently, the sample was rotated about the vertical axis with a step of 1$^\circ$ over a scanning range of 140$^\circ$. Inelastic neutron scattering data were collected using a PG(002) monochromator with an incident energy of $E_i = 9\ \text{meV}$, combined with a Fermi chopper rotating at 360~Hz using a counting time of 3.6 min/point.  Under this condition, the HYSPEC provides an energy resolution  better than  $100\mu$eV  at an energy transfer of
6 meV (Fig.~\ref{fig:Eresolution}). 

\begin{figure}[t]
    \centering
    \includegraphics[width=0.8\columnwidth]{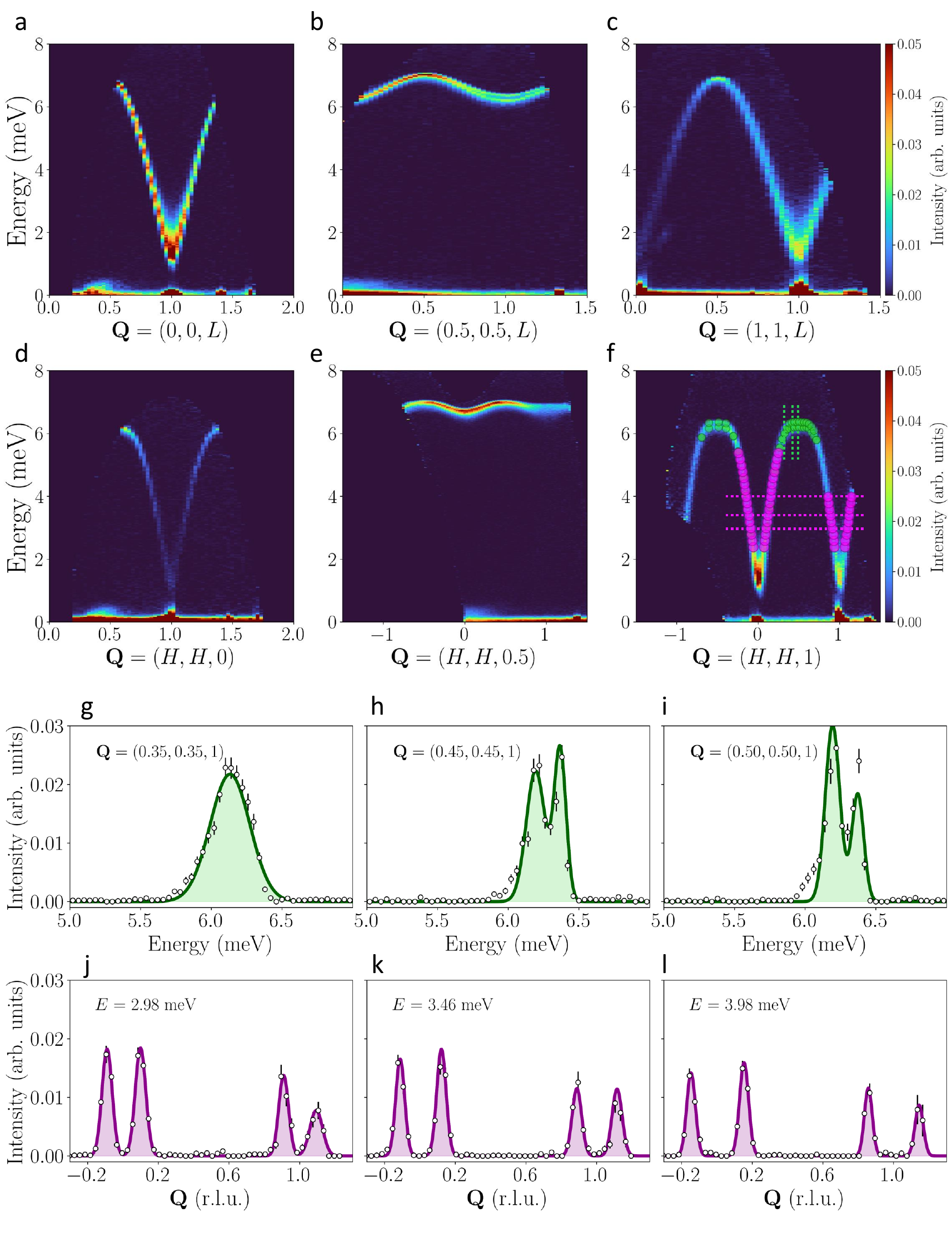}
    \caption{
\textbf{Unpolarized neutron scattering dispersion maps and fits.} (a--f) Dispersion spectra of MnF$_2$ measured with incident energy $E_i = 9$~meV along $\boldsymbol{Q} = (0,0,L), (0.5,0.5,L), (1,1,L), (H,H,0), (H,H,0.5)$, and $(H,H,1)$. Gaussian fits are shown as green and magenta circles in (f), corresponding to slices at fixed $\boldsymbol{Q}$ or energy $E$. Green vertical and magenta horizontal dashed lines indicate slices exemplified in (g--l) for $H = 0.35, 0.45, 0.50$~r.l.u. and $E = 2.98, 3.46, 3.98$~meV. Black markers show experimental integrated intensities. In (g--i), the solid green line shows the total fit from one or two Gaussians, with shaded green areas highlighting individual components. In (j--l), the solid magenta line shows the total fit from four Gaussians, with shaded pink areas highlighting the components.
}
    \label{fig:fits}
\end{figure}

The collected data were then used to extract the scattered intensity as a function of momentum \textbf{q},  and energy transfer $E$. Then, intensity dispersion maps of the dynamical structure factor $S(\boldsymbol{Q},E)$ were obtained from slices of the
four-dimensional data set using the SHIVER package of the MANTID software \cite{shiverref}. Dispersion maps were obtained using an integration window of 0.04 meV and 0.03~r.l.u. along the perpendicular wavevector direction, corresponding to 0.038 and 0.056 \AA$^{-1}$ along the $[HH0]$ and $[00L]$ directions, respectively. This corresponds to the estimated resolution of the instrument. We thus obtain the dispersions along $L$ for $H = K$ = 0.5, 0.8, 0.25, 0, and 1~r.l.u. and along $H,H$ for $L$ = 0, 0.5, 0.6, 0.75, 0.8, and 1~r.l.u. as well as the dispersion along $\boldsymbol{Q} = [H, H, H]$, providing a large data set, as shown by some relevant dispersion maps in Fig.~\ref{fig:fits} obtained along high-symmetry directions. For each of these dispersions, the data were fitted using a function composed of one or two Gaussian peaks depending on the case for fixed $\boldsymbol{Q}$ scattering vector slices and by up to four Gaussians for constant energy $E$ cuts. The uncertainties from the fit covariance were negligible compared to the intrinsic mode widths in $\boldsymbol{Q}$ and $E$, and were thus ignored. The intrinsic widths, expressed as full width at half maximum (FWHM) with $\mathrm{FWHM} \simeq 2.355\,\sigma$, were used as uncertainties to weight each point in the least squares minimization used to extract parameters of the Hamiltonian as described later. In total, over 1600 experimental data points were extracted to constrain the theoretical model.

\subsubsection*{Polarized Inelastic Scattering Measurements}

Polarized inelastic neutron scattering data were acquired in a half-polarized configuration, using a Heusler crystal to polarize the incident beam, without polarization analysis of the scattered neutrons. 
Measurements were performed with $E_i = 9\ \text{meV}$. As in the unpolarized neutron measurement, the $[1\bar{1}0]$ axis was oriented vertically, allowing measurements in the $(HHL)$ scattering plane. We collected data spanning a rocking scan angle of 58$^\circ$, with 1$^\circ$ step and using a counting time of 7 min/ point.

For an initially polarized neutron beam, the expression for the inelastic scattering intensity in the limit where there is an exact $U(1)$ symmetry in the magnon Hamiltonian can be written as~\cite{mcclarty2025}:
\[
I_n(\mathbf{Q},\mathbf{P}_{\rm in}) = \frac{1}{2}\left[1 + (\hat{\boldsymbol{Q}} \cdot \hat{\mathbf{N}})^2 + 2 (-)^n (\mathbf{P}_{\rm in} \cdot \hat{\boldsymbol{Q}})(\hat{\boldsymbol{Q}} \cdot \hat{\mathbf{N}})\right] C_{\mathbf{Q}}
\]
for mode $n$. This expression shows that the polarization-dependent component of the cross-section is maximized when the incident neutron polarization is aligned with both the Néel vector $\mathbf{N}$ and the scattering vector $\boldsymbol{Q}$. However, the chiral splitting, which arises from altermagnetism, appears only along specific symmetry directions. For our purposes, it is sufficient to probe along the $[HH1]$ direction where we expect maxima at the $\boldsymbol{Q} = [0.25, 0.25, 1]$ and $\boldsymbol{Q} = [0.75, 0.75, 1]$ points of the Brillouin zone, located within the $(HHL)$ scattering plane.

\begin{figure}[t]
    \centering
    \includegraphics[width=\columnwidth]{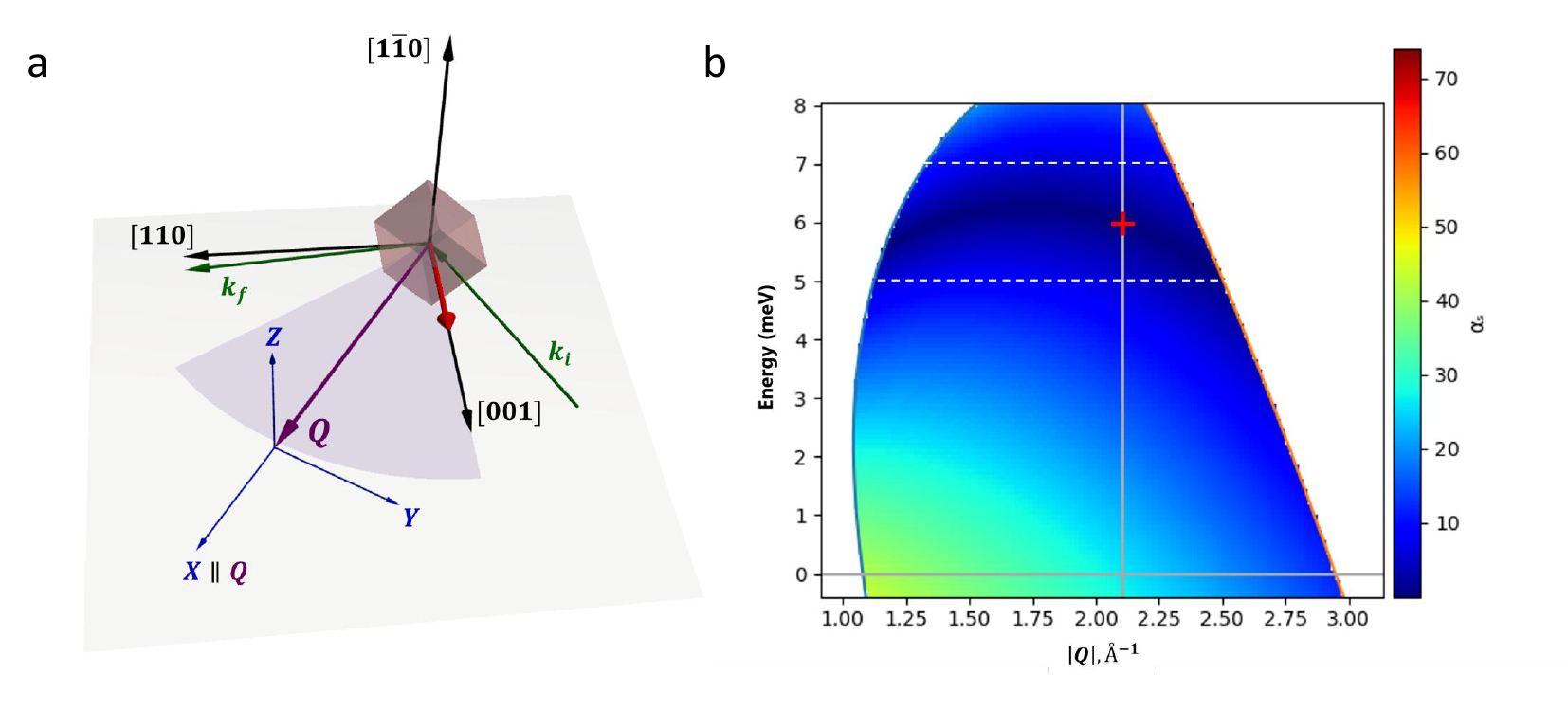}
    \caption{(a) Scattering geometry for $\boldsymbol{Q} = (1/2, 1/2, 1)$ (in purple) within the $(HHL)$ scattering plane. The $\boldsymbol{X}, \boldsymbol{Y}, \boldsymbol{Z}$ (blue vectors) with $\boldsymbol{X} \parallel \boldsymbol{Q}$, $\boldsymbol{Y} \perp \boldsymbol{Q}$ and $\boldsymbol{Z} \parallel [1 \bar{1} 0]$ corresponds to the Blume-Maleev frame. Our HYSPEC experiment was performed by maximizing the incident polarization $\mathbf{P}_{\rm in}$ with respect to $\boldsymbol{Q}$ at 6 meV energy transfer. The purple arc-circle denotes the $\boldsymbol{Q}$-range, i.e. between $\boldsymbol{Q} = (0, 0, 1)$ and $\boldsymbol{Q} = (1.25, 1.25, 1)$ where the polarization $\mathbf{P}_{\rm in}$ keeps at least $90$~\% of its maximum value within this $\boldsymbol{Q}$-range and 1-7 meV energy transfer. (b) Variation of the angle $\alpha_s$  between the $\bf \boldsymbol{X}$ incident polarization direction and the momentum transfer vector given in $|\boldsymbol{Q}|$ (\AA$^{-1}$) for a $9$ meV incident energy on HYSPEC. $\alpha_s=0$ at $\boldsymbol{Q}$=(0.5,0.5,1) and $E$ = 6 meV as indicated by the red cross. The dashed lines correspond to the region $(5-7)$ meV where the chiral splitting is seen on our polarized inelastic scattering data. In this region, $\alpha_s$ stays close to zero.} 
    \label{fig:Scharpf}
\end{figure}

To maximize the polarization-dependent component of the cross-section, the incident neutron polarization $\mathbf{P}_{\rm in}$ was optimized for $\boldsymbol{Q} = [0.5, 0.5, 1]$ and 6~meV energy transfer, which lies near both the scattering vector $\boldsymbol{Q}$ and the Néel vector $\mathbf{N}$. This orientation was achieved using the XYZ coil set of the HYSPEC instrument. 
This configuration allowed to keep $90$\% of the polarization of $\mathbf{P}_{\rm in}$ within the $[0, 0, 1] - [1.25, 1.25, 1]$ $\boldsymbol{Q}$ range and 1-7~meV energy transfer range as shown in Figure~\ref{fig:Scharpf}.

Measurements were performed consecutively with  $+\mathbf{P}_{\rm in}$ and $-\mathbf{P}_{\rm in}$ incident neutron polarizations. The reversal between $+\mathbf{P}_{\rm in}$ and $-\mathbf{P}_{\rm in}$ was achieved either by inverting the corresponding component of the magnetic field in the XYZ coil system or by flipping the neutron spin using the Mezei flipper.

Data reduction was performed using the \texttt{Shiver} software package \cite{shiverref}. The $S(\boldsymbol{Q}, \omega)$ maps shown in Figures~2 and 3 were generated with energy integration $\Delta E = 0.04\ \text{meV}$ and reciprocal space integration windows of $\Delta H = \Delta K = \Delta L = 0.03\ \text{r.l.u.}$

\subsubsection*{Evaluation of the chiral contribution to the total scattered intensity}

\begin{figure}[t]
    \centering
    \includegraphics[width=1\columnwidth]{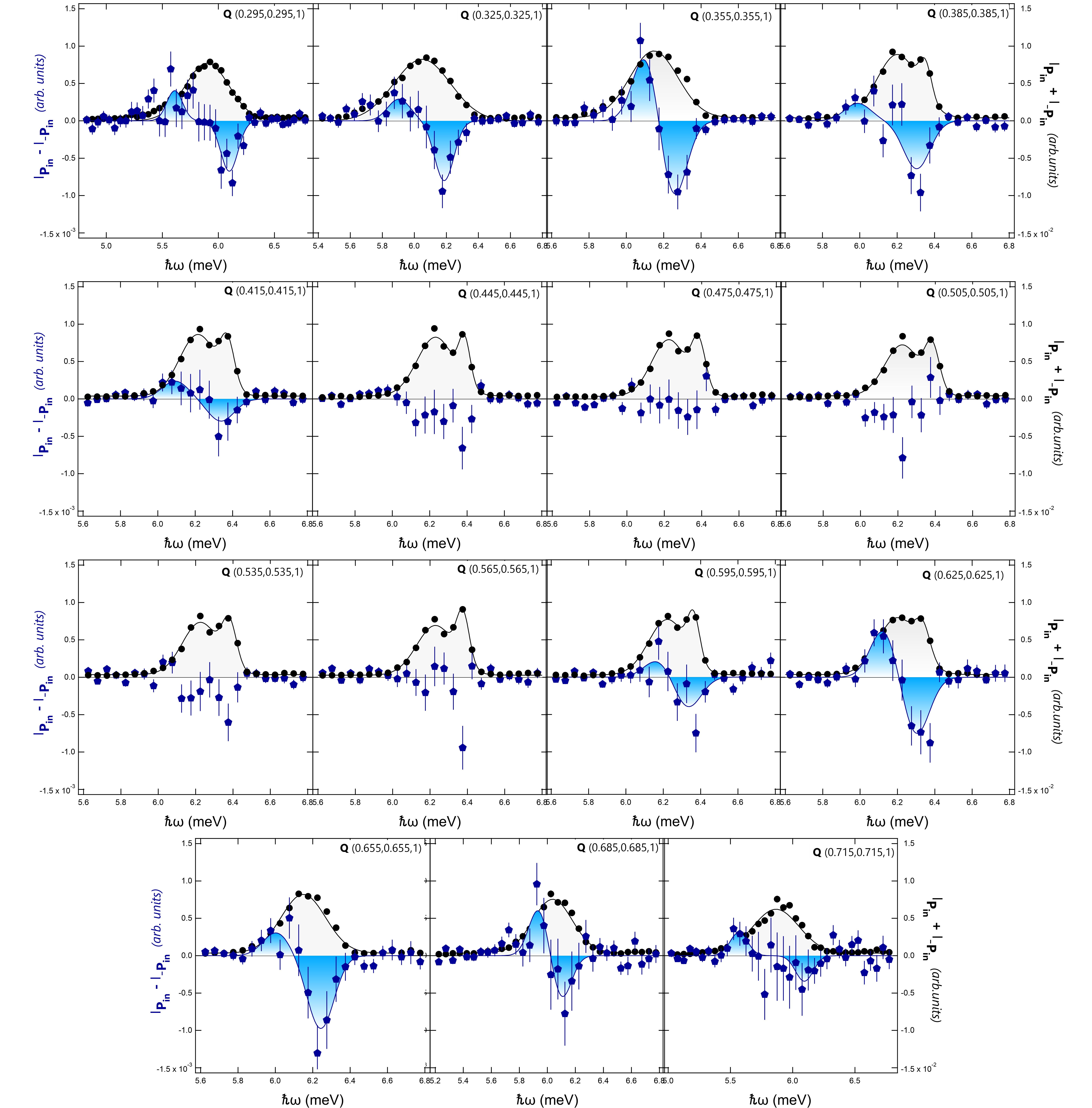}
    \caption{$\boldsymbol{Q}$ dependence of energy scans obtained  from experimental sum (black points) and difference (blue points) between $\mathbf{P}_{\rm in}$ and $\mathbf{P}_{\rm- in}$ cross sections. Black curves are guide to the eye. Blue curves are results from a double Gaussian fit. All data herein are integrated over 0.05~meV steps along energy and 0.03~r.l.u. steps in $\boldsymbol{Q}$ along the plotted axis. The transverse-$\boldsymbol{Q}$ averaging window is 0.1~r.l.u. in both cases.} 
    \label{chiral_qdep}
\end{figure}

Fig.~\ref{chiral_qdep} shows the $\boldsymbol{Q}$-dependence of the chiral contribution $I_{\mathbf{P}_{\rm in}}$-$I_{\mathbf{P}_{\rm -in}}$ to the total cross section $I_{\mathbf{P}_{\rm in}}$+$I_{\mathbf{P}_{\rm -in}}$. The data show that the chiral response vanishes at the zone boundary (around H=0.5) as expected from the chiral altermagnetic splitting of the magnon bands.

\begin{figure}[t]
    \centering
    \includegraphics[width=0.8\columnwidth]{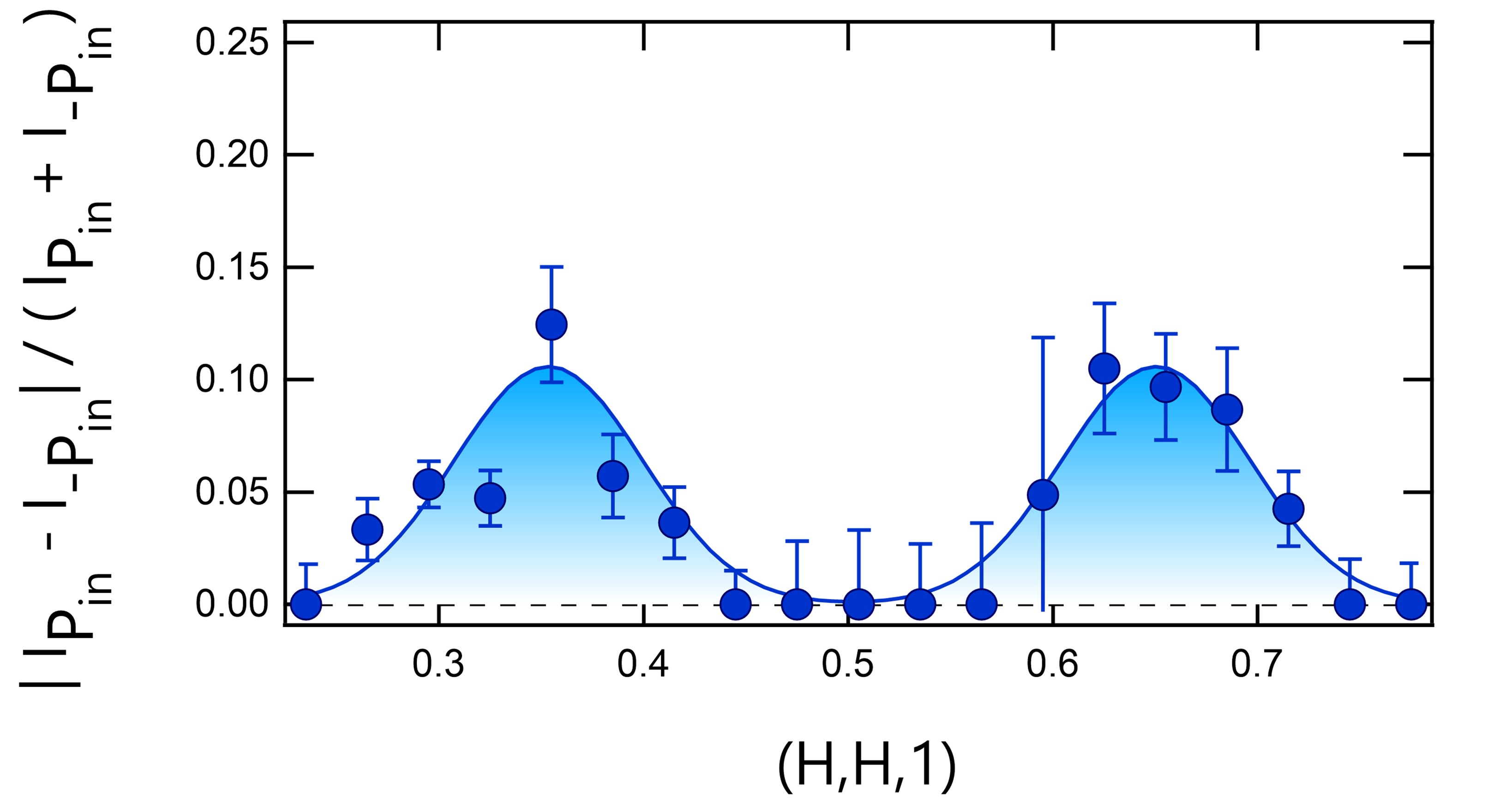}
    \caption{$\boldsymbol{Q}$ dependence of of the ratio between the integrated intensities of the chiral contribution $|\mathbf{P}_{\rm in}$ - $\mathbf{P}_{\rm- in}|$ and the total cross section $\mathbf{P}_{\rm in}$ + $\mathbf{P}_{\rm- in}$ corresponding to areas under Gaussian fits to the data shown in Fig.~\ref{chiral_qdep}. The blue curve is a guide to the eye.} 
    \label{chiralratio}
\end{figure}

Fig.~\ref{chiralratio} shows the resulting Q-dependence of the ratio between the integrated intensities of the chiral contribution and the total cross section. The chiral contribution is found to be maximal at  H=0.35 and 0.65, where it amounts to $\sim 11\pm 1 \%$ of the total cross section. 

\subsubsection*{Minimal Model}

\begin{figure}[t]
    \centering
    \includegraphics[width=.98\columnwidth]{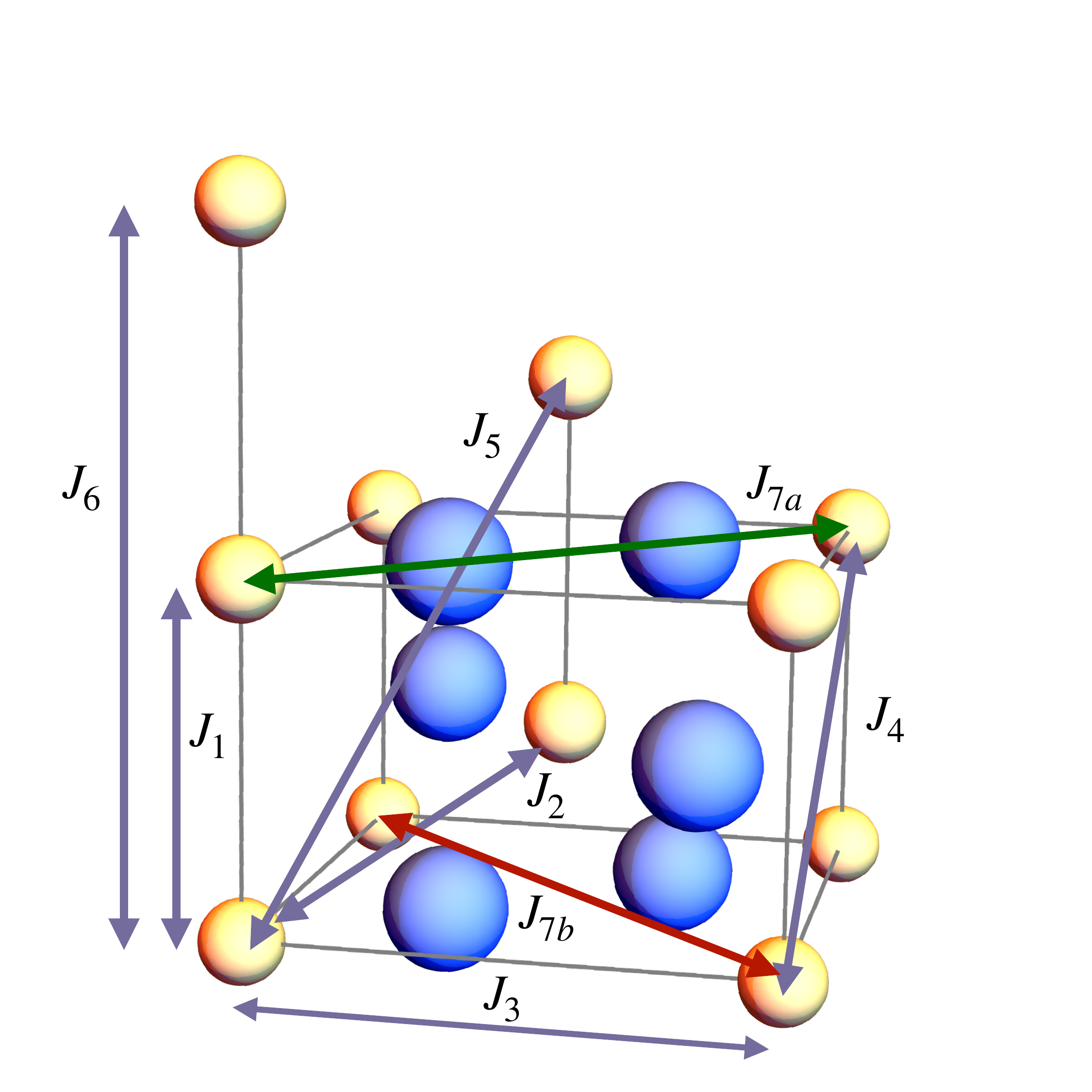}
    \caption{Figure showing the crystal structure of MnF$_2$ and $n$th nearest neighbor bonds used to specify the range of exchange couplings $J_n$.}
    \label{fig:crystal}
\end{figure}

The magnetic manganese ions have spin $S=5/2$ with no orbital component so the natural model to understand the magnetism in this material consists of Heisenberg exchange couplings and the long range dipolar coupling. Thus we consider Hamiltonian:
\beq
H = \sum_{\langle i,j\rangle_n} J_{n} \mathbf{S}_i \cdot \mathbf{S}_j + D \sum_{i,j} \frac{1}{\vert \mathbf{R}_{i,j} \vert^3} \left\{ \mathbf{S}_i\cdot\mathbf{S}_j - 3  \left( \mathbf{S}_i\cdot \hat{\mathbf{R}}_{i,j} \right)  \left( \mathbf{S}_j\cdot \hat{\mathbf{R}}_{i,j} \right)     \right\}
\label{eq:model}
\eeq
acting between the local moments on the $(0,0,0)$ and $(1/2,1/2,1/2)$ positions of the tetragonal primitive cell with $c/a=0.679$. Figure~\ref{fig:crystal} illustrates the symmetry inequivalent pairs coupled by the nth parameters $J_n$. 

The manganese ions carry spin $S=5/2$ and magnetic moment $g_S S \mu_{\rm B}$. The scale of the long-range magneto-static dipolar coupling is
\beq
D = \mu_0 \frac{\mu^2 \mu_{\rm B}^2 }{4\pi R_{\rm nn}^3} = 0.0518 \ {\rm meV}
\eeq
where we have used lattice parameters $a=4.87\AA$ and $c=3.31 \AA$ such that the nearest neighbour distance $R_{\rm nn}$ is along the $c$ axis: $c=3.31 \AA$.

It is straightforward to see that each of the couplings $J_{n}$ for $n=1,\ldots,6$ has a higher symmetry than the space group of MnF$_2$. In particular, a model with any of these Heisenberg couplings has a translation symmetry that connects the two magnetic sublattices implying that these are simple antiferromagnets with no altermagnetic splitting. The seventh neighbor bonds are special because they split into two classes that cannot be connected under the symmetries of the crystal. We denote the two couplings at the seventh neighbor $J_{7a}$ and $J_{7b}$. When these couplings satisfy $J_{7a}\neq J_{7b}$ the model has the precisely the symmetries of the MnF$_2$ crystal. This means that these are the shortest range Heisenberg couplings that can lead to altermagnetism. This phenomenon of symmetry enhancement and emergence of altermagnetism via Heisenberg couplings at a given shell number is detailed in Ref.~\cite{Gohlke2023,dagnino2024} for all crystal structures.

A similar argument shows that the long-range dipolar coupling alone cannot lead to altermagnetism. In particular, the dipolar coupling couples the Mn sites which lie on a body-centred tetragonal lattice which, again, links the magnetic sublattices by a translation and time reversal. The ``decoration" of the lattice with F ions breaks down this symmetry to $C_{4z}$, translation and time reversal but the dipolar coupling in Eq.~\ref{eq:model} has a higher symmetry. This dipolar coupling is an approximation to reality as it acts between point-like moments (neglecting higher order magnetic multipoles) when, in fact, the dipolar coupling depends on the local magnetization density. In order to build the correct symmetries into the model we may consider anisotropic short-range couplings $-$ we consider these in the following section.  

\subsubsection*{Spin Wave Theory and Neutron Cross Section}

The calculations of the inelastic neutron scattering intensity were carried out using a multi-boson (or flavour wave) method on the $S=5/2$ states of MnF$_2$. We first carry out self-consistent local mean field theory on the model in equation~\ref{eq:model}  to obtain the spectrum on each magnetically distinct site $\vert i,\mu\rangle$. The minimal coupling that leads to a simple antiferromagnet is $J_2>0$. This antiferromagnetic interaction is entirely isotropic but, in reality, the magnetic structure of MnF$_2$ has moments oriented parallel to the c axis and the spin wave spectrum is gapped. The long-range dipolar coupling introduces precisely this kind of easy axis anisotropy.  

We introduce boson operators that create the local states on top of the vacuum $\vert i,\mu\rangle \equiv A^\dagger_{i,\mu} \vert {0} \rangle$ where $i$ labels the site and $\mu$ the state running from $0$ to $2S$. There is a constraint that the boson number is equal to one per site $\sum_\mu A^\dagger_{i\mu}A_{i\mu}=1$. For the purposes of formulating a spin wave expansion we take 
\beq
\sum_\mu A^\dagger_{i\mu}A_{i\mu}=M
\eeq
for fixed $M$. This plays the role of $S$ in usual spin wave theory. In addition, the ground state is to be regarded as a condensate of bosons so that
\beq
A_{i0} = A^\dagger_{i0} = \sqrt{ M - \sum_{\mu=1}^{2S} A^\dagger_{i\mu}A_{i\mu}  }.
\eeq
We now write the operators in the Hamiltonian in the on-site basis
\beq
\hat{S}^\alpha_i = \sum_{\mu,\nu} \vert i\mu \rangle \langle i\mu \vert  \hat{S}^\alpha_i \vert i\nu \rangle \langle i\nu \vert \rightarrow \sum_{\mu,\nu} \langle i\mu \vert  \hat{S}^\alpha_i \vert i\nu \rangle A^\dagger_{i\mu}A_{i\nu}.
\eeq
Now we expand the Hamiltonian around the mean field ground state. We also redefine $B_{i}^\alpha = M \tilde{B}_{i}^\alpha$.  We find
\beq
H = M^2 H^{(0)} + M^{3/2} H^{(1)} + M H^{(2)} + \ldots 
\eeq
where 
\beq
H^{(0)} =  \frac{1}{2}\sum_{i,j} J_{ij}^{\alpha\beta} [\hat{S}^\alpha_i]_{00} [ \hat{S}^\beta_j ]_{00} - \sum_{i,\alpha} B_i^\alpha [ \hat{S}^\alpha_i ]_{00}
\eeq
using notation $ \langle i\mu \vert  \hat{S}^\alpha_i \vert i\nu \rangle\equiv [\hat{S}^\alpha_i]_{\mu\nu}$. 

For the single boson terms we find 
\begin{align}
H^{(1)} = A_{i\nu} \left[ \sum_j \left( J_{ij}^{\alpha\beta} [ \hat{S}_i^\alpha ]_{0\nu}  [ \hat{S}_j^\beta ]_{00} \right) -  \tilde{B}_{i}^\alpha [ \hat{S}_i^\alpha ]_{0\nu}   \right] \nonumber \\
+ A^\dagger_{i\mu} \left[ \sum_j \left( J_{ij}^{\alpha\beta} [ \hat{S}_i^\alpha ]_{\mu 0}  [ \hat{S}_j^\beta ]_{00} \right) -  \tilde{B}_{i}^\alpha [ \hat{S}_i^\alpha ]_{\mu 0}   \right] 
\end{align}
These should vanish in any local minimum solution of the self-consistent mean field equations.

Finally, for the quadratic Hamiltonian we use the notation
\beq
H^{(2)} = \frac{1}{2} \sum_{\mathbf{k}} \boldsymbol{\Upsilon}_{\mathbf{k}}^\dagger \left( \begin{array}{cc} 
\boldsymbol{A}(\mathbf{k}) &  \boldsymbol{B}(\mathbf{k})  \\
\boldsymbol{B}^\star(-\mathbf{k}) &  \boldsymbol{A}^*(-\mathbf{k})  
 \end{array} \right) \boldsymbol{\Upsilon}_{\mathbf{k}} \equiv 
 \frac{1}{2} \sum_{\mathbf{k}} \boldsymbol{\Upsilon}_{\mathbf{k}}^\dagger \mathbf{M}_{\mathbf{k}}\boldsymbol{\Upsilon}_{\mathbf{k}}
\eeq
with $\boldsymbol{\Upsilon}_{\mathbf{k}} = (A_{\mathbf{k}a1}, \ldots , A_{\mathbf{k}b2S},A^\dagger_{-\mathbf{k}a1},\ldots A^\dagger_{-\mathbf{k}b2S})^T$ where $a$ and $b$ label the two magnetic sublattices. Here,
\begin{align}
A_{ab}^{\mu\nu}(\mathbf{k}) & = \sum_{\alpha\beta} J_{ab}^{\alpha\beta}(\mathbf{k}) [ \hat{S}_a^\alpha ]_{\mu 0} [ \hat{S}_b^\beta ]_{0\nu} + \delta_{ab} \sum_c J_{ac}^{\alpha\beta}(\mathbf{k}=0) \left[   [ \hat{S}_a^\alpha ]_{\mu \nu} [ \hat{S}_c^\beta ]_{0 0} -  \delta_{\mu\nu} [ \hat{S}_a^\alpha ]_{0 0} [ \hat{S}_c^\beta ]_{00}  \right] \nonumber \\
& - \sum_{\alpha} \tilde{B}_a^\alpha \left(  [ \hat{S}_a^\alpha ]_{\mu\nu} - \delta_{\mu\nu} [  \hat{S}_a^\alpha ]_{00} \right)
\end{align}
and
\begin{align}
B_{ab}^{\mu\nu}(\mathbf{k}) & = \sum_{\alpha\beta} J_{ab}^{\alpha\beta}(\mathbf{k}) [ \hat{S}_a^\alpha ]_{\mu 0} [ \hat{S}_b^\beta ]_{\nu 0} 
\end{align}

The contribution of the long-ranged dipolar coupling to $J_{ab}^{\alpha\beta}(\mathbf{k})$  is computed via an Ewald summation. 

The above formulation is general enough to be applicable to more general Hamiltonians $-$ for example those involving Stevens operators as single ion terms. 

To find the spectrum we diagonalize the 
$8S\times 8S$ quadratic Hamiltonian via a bosonic Bogoliubov transformation. Introducing  $\boldsymbol{\eta} = {\rm diag}(1,\ldots,1,-1,\ldots,-1)$ consisting of $4S$ ones then $4S$ minus ones along the diagonal, we find a $\mathbf{T}_{\mathbf{k}}$ such that 
\beq
\mathbf{T}_{\mathbf{k}} \boldsymbol{\eta} \mathbf{M}_{\mathbf{k}} \mathbf{T}^{-1}_{\mathbf{k}} =\left( \begin{array}{cc} \boldsymbol{\epsilon}_{\mathbf{k}} & 0 \\ 0 & -\boldsymbol{\epsilon}_{\mathbf{k}} \end{array} \right)
\eeq
where the $10$ mode energies are organized as $\boldsymbol{\epsilon}_{\mathbf{k}}={\rm diag}(\epsilon_{1\mathbf{k}},\ldots,\epsilon_{4S\mathbf{k}})$. In order to preserve the bosonic commutation relations the diagonalizing transformation must satisfy $\mathbf{T}_{\mathbf{k}}^{-1} = \boldsymbol{\eta} \mathbf{T}_{\mathbf{k}}^{\dagger} \boldsymbol{\eta}$. 

\subsubsection*{Polarized neutron cross section}

Inelastic neutron scattering on magnetic materials is most often carried out without regard to the spin polarization of the neutron. Nevertheless there are many instruments that both allow  the in-going neutron polarization to be controlled and for the polarization of the out-going neutron to be measured. The HYSPEC experiment was operated in a half-polarized mode $-$ measuring the total out-going intensity for a fixed initial polarization. The inelastic neutron scattering cross section for energy transfer $\omega$ and momentum transfer $\boldsymbol{Q}$ for this setting is 
\beq
\left( \frac{d^2\sigma}{ d\Omega d\omega } \right) \propto \int dt e^{-i\omega t} \left[ \langle \mathbf{M}_{-\boldsymbol{Q}}^{\perp} \cdot  \mathbf{M}_{\boldsymbol{Q}}^{\perp}(t)  \rangle + i \mathbf{P}_{\rm in} \cdot  \langle \mathbf{M}_{-\boldsymbol{Q}}^{\perp} \times  \mathbf{M}_{\boldsymbol{Q}}^{\perp}(t)  \rangle  \right]
\label{eq:crosssection}
\eeq
where $\mathbf{M}_{\boldsymbol{Q}} = \sum_{\mathbf{r},n} \mu_n f_n(\boldsymbol{Q}) e^{i\boldsymbol{Q}\cdot (\mathbf{r}+\boldsymbol{\delta}_n )} \mathbf{S}_{\mathbf{r},n}$ and $ \mathbf{M}_{\boldsymbol{Q}}^{\perp}\equiv \hat{\boldsymbol{Q}} \times \left( \mathbf{M}_{\boldsymbol{Q}} \times  \hat{\boldsymbol{Q}} \right)$. Here $\mu_n$ are the moments on sublattice $n$ which have form factor $f_n(\mathbf{k})$.  The initial neutron polarization $\mathbf{P}_{\rm in}$ was chosen to align or anti-align with the crystal $[1/2~1/2~1]$ direction (and therefore varied with respect to the lab frame). 

The experiment was first carried out with $\mathbf{P}_{\rm in}$ fixed and then repeated with the sign reversed polarization. Evidently, by subtracting the two intensities we isolate the contribution coming from the second term on the right-hand-side of Eq.~\ref{eq:crosssection} while the sum of the two intensities is the usual unpolarized inelastic cross section. The difference map is sensitive to the chirality of the magnon modes as we shall see. 

We compute the contribution from each term within the linear flavour wave theory using
\beq
\hat{S}_i^\alpha \rightarrow M\langle i,0 \vert S_i^\alpha \vert i,0 \rangle + \sqrt{M} \sum_{p=1}^{2S} \left( A_{ip}^\dagger [ \hat{S}_i^\alpha ]_{p0}  + A_{ip} [ \hat{S}_i^\alpha ]_{0p} \right).
\eeq
The correlation functions corresponding to the first and second terms on the right-hand-side of Eq.~\ref{eq:crosssection} are, respectively
\begin{align}
& \left( \frac{d^2\sigma}{ d\Omega d\omega } \right)_{\rm un-pol} \propto M f^2_{\rm Mn}(\boldsymbol{Q}) \sum_{ab, \alpha\beta = x,y}  \left( \delta_{\alpha\beta} - \hat{Q}_\alpha \hat{Q}_\beta \right) \langle \hat{S}^{\alpha}_{-\boldsymbol{Q},a,-\omega} \hat{S}^\beta_{\boldsymbol{Q},b,\omega} \rangle  \\
& \left( \frac{d^2\sigma}{ d\Omega d\omega } \right)_{\rm pol} \propto i M f^2_{\rm Mn}(\boldsymbol{Q})\left(\mathbf{P}_{\rm in}\cdot \hat{\boldsymbol{Q}} \right) \hat{Q}^z \epsilon^{z\alpha\beta} \sum_{ab} \langle \hat{S}^{\alpha}_{-\boldsymbol{Q},a,-\omega} \hat{S}^\beta_{\boldsymbol{Q},b,\omega}\rangle
\end{align}

For Mn$^{2+}$ the magnetic form factor is $f_{\rm Mn}(\mathbf{k}) = \left\langle j_0(\mathbf{k}) \right\rangle$ approximated by \cite{allen1992international}
\beq
f_{\rm Mn}(s) = A\exp(-as^2)+B\exp(-bs^2)+C\exp(-cs^2) + D
\eeq
with $s= \sin\theta/\lambda$ in $\AA$ and parameters in Table~\ref{tab:ff}.
\begin{table}[ht]
\caption{Table of form factor coefficients.}
\begin{center}
\begin{tabular}{|| c c c c c c c ||} 
 \hline
 A & a & B & b & C & c & D\\ [0.5ex] 
 \hline\hline
 0.4220 & 17.6840 & 0.5948 & 6.0050 & 0.0043 & -0.6090 & -0.0219 \\
 \hline 
\end{tabular}
\end{center}
\label{tab:ff}
\end{table}

\subsubsection*{Parameterization of the dispersion relations}

\begin{figure}[!]
\includegraphics[width=\linewidth]{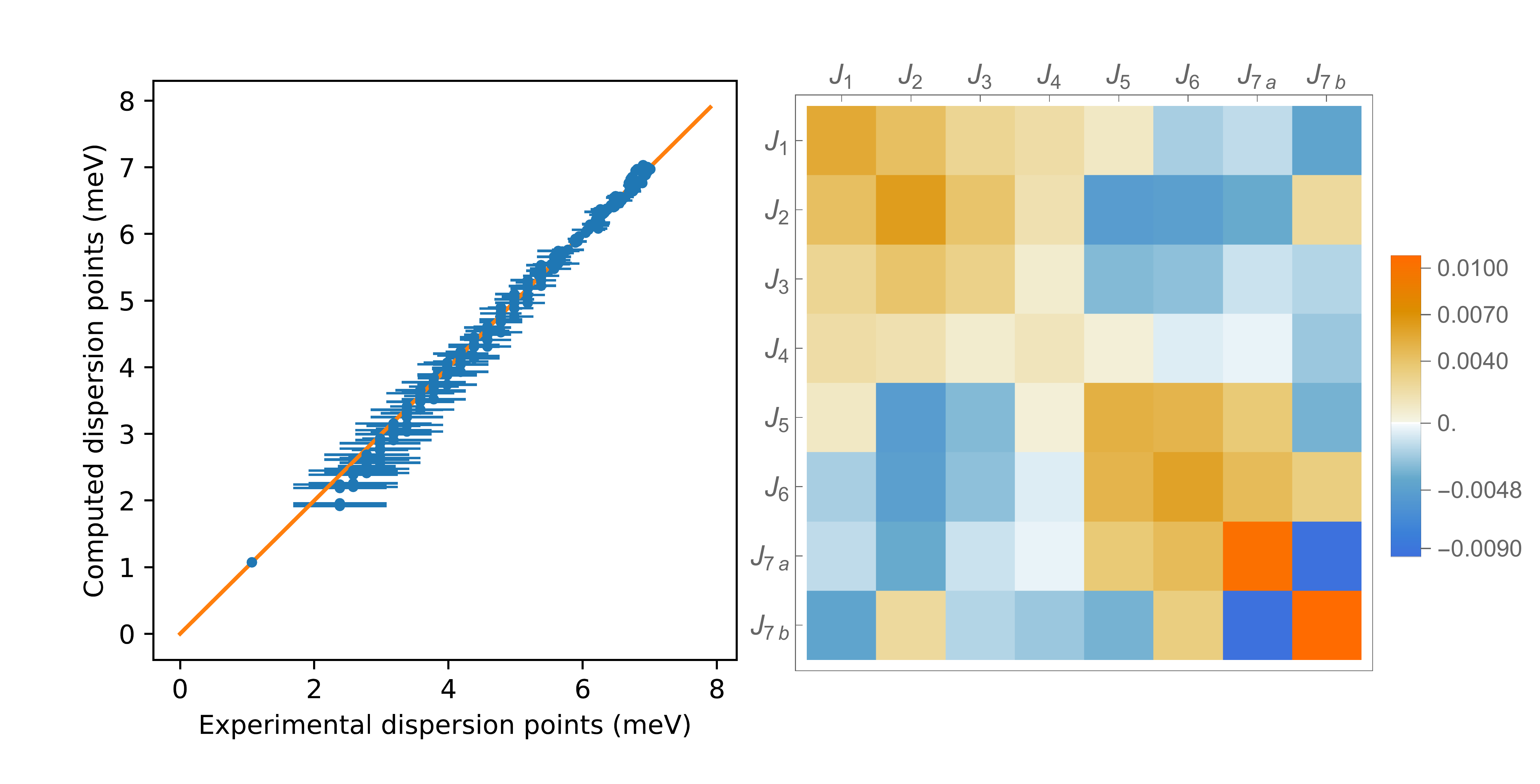}
\caption{(Left) Plot of the dispersion points extracted from the scattering data with their associated uncertainty and the corresponding points calculated from the best fit exchange model. (Right) From the distributions of the eight fitted exchange parameters extracted from the experimental data including noise, the covariance matrix $M_{ab}$ is computed. The plot shows ${\rm sgn}(M_{ab})\sqrt{\vert M_{ab}\vert}$.}
\label{fig:compare}
\end{figure}

The minimal model with Heisenberg exchange couplings $J_1.\ldots,J_{7a}, J_{7b}$ with fixed dipolar coupling was used to parameterize dispersion relations extracted from the data. To obtain the model parameters, a least squares minimization was carried out using a Levenberg-Marquardt algorithm implemented in Julia. In other words, given a set of dispersion points from the scattering data $\omega^{(e)}_n$ within energy uncertainty $\sigma_n$ and computed points at the same wavevectors $\omega^{(c)}_n$, the expression
\beq
S = \sum_n \left( \frac{\omega^{(e)}_n - \omega^{(c)}_n}{ \sigma^{(c)}_n } \right)^2
\eeq
was minimized. The experimental dispersion points and those computed from the best fit are shown in Fig.~\ref{fig:compare}(left panel). Error bars were extracted from the experimental data as discussed above. Clearly the fit is good within the experimental uncertainties. The plots in the main text were computed using these parameters. 

\begin{figure}[!]
\begin{center}
\includegraphics[width=0.6\linewidth]{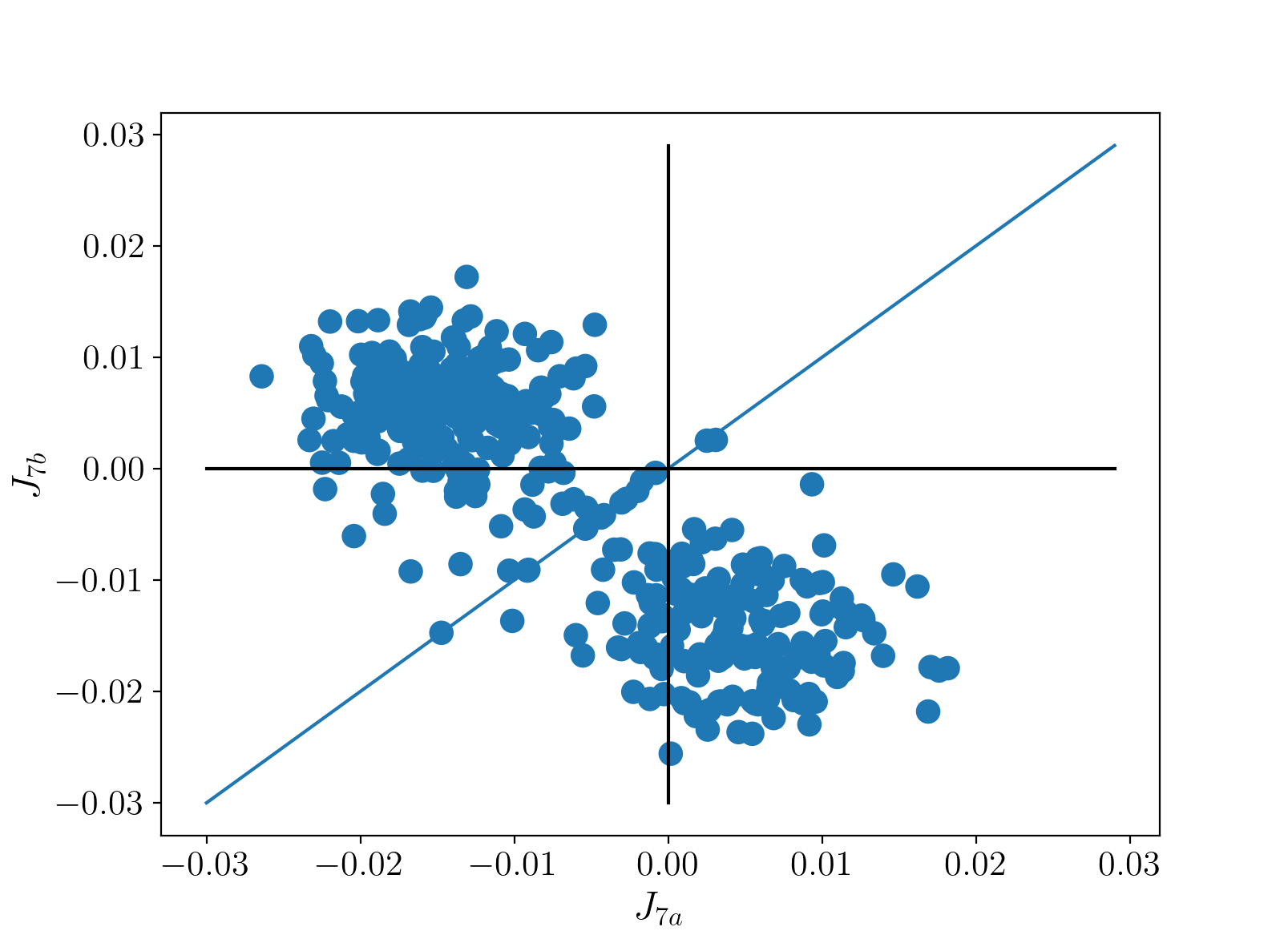}
\caption{Plot showing $J_{7a}$ and $J_{7b}$ from optimized set of parameters to noisy data as described in the main text. The distribution is bimodal with an approximate symmetry interchanging the couplings.}
\end{center}
\label{fig:j7}
\end{figure}

In order to estimate uncertainties in the exchange couplings, we generated data from the original dispersions by adding a Gaussian random variable to each point with mean zero and standard deviation extracted from the FWHM of the fitted intensity peaks. We carried out an optimization of the couplings for many noise realizations. The distribution of each coupling $J_n$ ($n\leq 6$) obtained in this way is peaked around some value and the quoted error below is the standard deviation from a Gaussian fit of that distribution. This analysis reveals that $J_4$ and $J_6$ are zero within errors. These couple identical magnetic sublattices. Within sampling errors the distribution of $J_{7a},J_{7b}$ is bimodal with reflection symmetry about $J_{7a}=J_{7b}$ indicating that the dispersion relations are insensitive to the relative sign of the difference of the couplings (Fig.~\ref{fig:j7}). Fig.~\ref{fig:compare}(right panel) illustrates the covariance matrix of the parameters extracted from fits to the noisy data. This shows strong correlations between several pairs of parameters and, in particular, highlights the anticorrelation between $J_{7a},J_{7b}$. 

\begin{figure}[!]
\begin{center}
\includegraphics[width=0.6\linewidth]{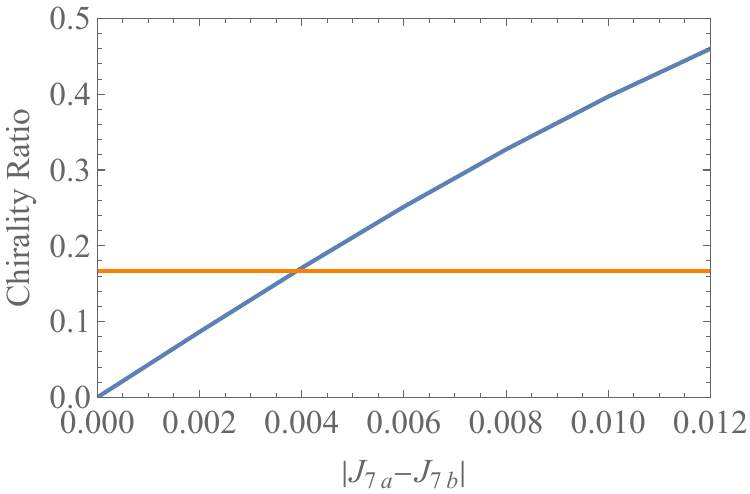}
\caption{Plot showing the calculated ratio of the chiral term in the neutron cross section at $[HH1]$ with $H=0.7$ and the total intensity for different $\vert J_{7a}-J_{7b}\vert$ with all other couplings fixed to their values in Eq.~\ref{eq:fitparameters}. The horizontal line indicates a value compatible with experiment having adjusted for the measured domain ratio. This constrains the magnitude of the altermagnetic splitting in MnF$_2$.  
}
\end{center}
\label{fig:chiralratio}
\end{figure}

We may further constrain $J_{7a}$ and $J_{7b}$ from the ratio of the chiral term to the total intensity extracted from the experiment. This ratio was found to be between $5\%$ and $11\%$ depending on the cut. Taking the ratio to be $10\%$ for the reported $4:1$ domain ratio, the ratio for a mono-domain sample would be about $16\%$. The figure~\ref{fig:chiralratio} constrains $\vert J_{7a}-J_{7b}\vert \approx 0.04$. We therefore take $J_{7a} = -0.006$ and $J_{7b} = -0.002$ about the center of mass values from the fit. 

The resulting exchange parameters from the distribution of best fit couplings are:
\begin{align}
& J_1 = -0.075(2) {\rm meV} \hspace{0.5cm} J_2 = 0.287(3)   {\rm meV} \hspace{0.5cm}  J_3 = -0.012(1)  {\rm meV} \nonumber \\
& J_4 = -0.001(1) {\rm meV} \hspace{0.5cm} J_5 = 0.008(2)  {\rm meV} \hspace{0.5cm}  J_6 = 0.001(2)  {\rm meV} \nonumber \\
& J_{7a} = -0.006(3) {\rm meV} \hspace{0.5cm} J_{7b} = -0.002(3)  {\rm meV} 
\label{eq:fitparameters}
\end{align}

For comparison, Nikotin et al. \cite{nikotin1969}, based on their data taken at the DR3 reactor facility in Denmark, carried out a fit of $J_1$, $J_2$ and $J_3$ with fixed dipolar coupling obtaining:
\beq
J_1 = -0.056(2) {\rm meV} \hspace{0.5cm} J_2 = 0.304(2) {\rm meV} \hspace{0.5cm}  J_3 = 0.008(2) {\rm meV}
\eeq
Compared to the figures directly reported in their paper, these use the convention with antiferromagnetic couplings positive and the couplings are multiplied by two to coincide with our Hamitonian convention.  

A much more recent study based on data from the CAMEA instrument at PSI in Switzerland uses the first three neighbor couplings and a single ion Ising anisotropic coupling $D_{c}$ instead of the long-range dipolar coupling \cite{morano2024absence}. With this Hamiltonian, their data is fit well using
\beq
J_1 = -0.0677(9) {\rm meV} \hspace{0.5cm} J_2 = 0.3022(6)  {\rm meV} \hspace{0.5cm}  J_3 = -0.0044(4)04  {\rm meV} 
\eeq
and $D_c = -0.0267(6)$ meV.

\subsubsection*{{\it A priori} constraints on the anisotropic couplings}
 
In this section, we consider possible origins of the measured polarization dependence of the magnon modes beyond the Heisenberg model considered in the main text. As we have discussed extensively, the principal couplings in MnF$_2$ are Heisenberg exchange as appropriate to pure spin $S=5/2$ and the long-range magnetostatic dipolar interaction. The dipolar coupling cannot introduce a chiral splitting because the coupling has the symmetries of the body-centered tetragonal lattice whereas the altermagnetism comes from the full MnF$_2$ crystal structure. For example, the $C_{4z}$ symmetry of the former is broken in the latter structure. Here we consider couplings allowed by the crystalline symmetries. 

With these observations in mind, we consider single ion terms and general two-spin interactions. It is also conceivable that $2n$-spin ($n\geq 2$) couplings play a role in the altermagnetism in MnF$_2$ but we leave an examination of these to future studies. 

\subsubsection*{Single ion couplings}

The local site symmetry is $D_{2h}$ that constrains the crystal field Hamiltonian such that it admits  
\beq
B_{2}^{0}, \hspace{4pt} B_{2}^{2}, \hspace{4pt} B_{4}^{0}, \hspace{4pt} B_{4}^{2}, \hspace{4pt} B_{4}^{4}. 
\eeq
Higher order couplings need not be considered within the $S=5/2$ multiplet. We note that the body-centred tetragonal lattice $-$ where only the manganese ions are present $-$ has a significantly higher symmetry  I4/mmm (\# 139) with $D_{4h}$ site symmetry allowing only 
\beq
B_{2}^{0} \hspace{4pt} B_{4}^{0} \hspace{4pt} B_{4}^{4}. 
\eeq
The fits presented in this article with fixed dipolar coupling capture the anisotropy gap very well so the $B_{2}^{0}\propto (S^z)^2 + {\rm const.}$ may be neglected consistent with spin-orbital effects being small. When the dipolar splitting is not resolved, the magnon dispersion relations are well described by Heisenberg exchange including further neighbour couplings with, in addition, a non-vanishing $B^2_0$ \cite{morano2024absence}. 

The principal interest here is in crystal field terms that are not admitted by the body-centred tetragonal lattice as these have symmetries that underlie altermagnetism. The simplest such coupling is $O_2^2 = S^x S^y + S^y S^x$. This operator, like most anisotropic couplings, breaks the U(1) symmetry of the Heisenberg couplings. With coupling $B_2^2$ on one sublattice, the $C_{4z}$ translation leads to $-B_2^2$ on the other sublattice. 

Now we consider an interacting spin model with Heisenberg couplings and long-range dipolar couplings in the presence of a small $B_2^2$ single-ion term. The mean field ground state for $B_2^2=0$ has moments along $\pm \hat{\mathbf{z}}$. When $B_2^2$ is switched on, there is a finite threshold before expectation values of $x-y$ spin components become nonzero in the ground state. Above the threshold in $B_2^2$ there is canting of the moments and a spin flop into a state with the moment entirely in the $x-y$ plane. The chirality is zero below the threshold. Only above the threshold may the chirality be nonzero. 

\subsubsection*{Two-spin couplings}

Turning now to two-spin terms out to the seventh neighbor, we compute the constraints coming from the crystal symmetries with the following results. For a given bond labelled by the sites at the endpoints $i,j$, we compute the form of the symmetry-allowed anisotropic exchange $J_{ij}^{\alpha\beta} S^\alpha_i S^\beta_j$ where $\alpha,\beta$ label the spin components. As a shorthand we use the notation $\mathbf{J}_n$ for the exchange couplings at the $n$th nearest neighbor bond referring to Fig.~\ref{fig:crystal}. 

For bonds connecting identical magnetic sublattices we find:
\[
\begin{array}{c@{\hspace{30pt}}c@{\hspace{30pt}}c@{\hspace{30pt}}c} 
\text{Bond J$_1 $} & \text{Bond J$_3$} & \text{Bond J$_4$} \\[5pt]
\begin{bmatrix}
xx & xy & 0 \\
xy & xx & 0 \\
0 & 0 & zz
\end{bmatrix} &

\begin{bmatrix}
xx & xy & 0 \\
xy & yy & 0 \\
0 & 0 & zz
\end{bmatrix} &
\begin{bmatrix}
xx & xy & xz \\
xy & yy & yz \\
xz & yz & zz
\end{bmatrix} \\[30pt]

\text{Bond J$_6$} & \text{Bond J$_7$} \\[5pt]
\begin{bmatrix}
xx & xy & 0 \\
xy & xx & 0 \\
0 & 0 & zz
\end{bmatrix} &
\begin{bmatrix}
xx & xy & 0 \\
xy & xx & 0 \\
0 & 0 & zz
\end{bmatrix}.
\end{array}
\]
For example, for $\mathbf{J}_1$, there are three allowed couplings: XXZ couplings and a symmetric off-diagonal exchange coupling. The $J_4$ bond has the lowest symmetry with $6$ allowed couplings. All of these bonds have only symmetric exchange. For exchange between magnetic ions on different sublattices we find:
\[
\begin{array}{c@{\hspace{30pt}}c@{\hspace{30pt}}c@{\hspace{30pt}}c} 
\text{Bond J$_2$} & \text{Bond J$_5$} \\[5pt]
\begin{bmatrix}
xx & xy & xz \\
xy & xx & -xz \\
zx & -zx & zz
\end{bmatrix} &
\begin{bmatrix}
xx & xy & xz \\
xy & xx & -xz \\
zx & -zx & zz
\end{bmatrix} \\[30pt]

\end{array}
\]
Now we find allowed DMI couplings $S^x_iS^z_j-S^z_iS^x_j$ and for $y,z$ components also.

We note that $J^{xz}$, $J^{yz}$, $J^{zx}$ and $J^{zy}$ terms do not enter the linear spin wave theory. Their effects emerge at higher order in perturbation theory through other further neighbor couplings. This may include symmetric off-diagonal exchange $J^{xy} = J^{yx}$ which is allowed for all neighbors at least out to $7$th neighbor. 

It is helpful to inspect the couplings that are allowed on the body-centered tetragonal lattice that is obtained from the structure we have been considering by removing the F ions. This structure does not admit altermagnetism. We find that the $J^{xy}$ coupling is forbidden on bonds $1,3,4$ and $6$ connecting identical magnetic sublattices. The seventh neighbor bonds $a$ and $b$ are identical once the fluoride decoration has been removed. On bonds $2$ and $5$, $J^{xy}$ remains but there is an additional constraint on the couplings given above of the form $J^{xz}=J^{zx}$. 

Comparing the couplings allowed for the rutile structure and the body-centered tetragonal lattice directly imply that the $J^{xy}$ coupling between different magnetic sublattices cannot lead to a net chirality. They do, however, lead to a splitting of the magnon bands as one might expect as they form a component of the dipolar coupling. Direct calculation of the effect of this coupling reveals further interesting features: although the coupling breaks the $U(1)$ symmetry of the Heisenberg model there is an accidental $U(1)^*$ symmetry of the linear spin wave Hamiltonian such that Goldstone modes are present in the spectrum. The coupling produces a splitting everywhere but the $(H0L)$ planes and the chiral term in the neutron cross section vanishes. The latter result may also be inferred from the argument of Ref.~\cite{mcclarty2025} where a general two-sublattice $U(1)$ breaking perturbation {\it as the sole cause of magnon splitting}
\beq
\delta \mathbf{M}_{\vec{k}} = 
  \left(
    \begin{array}{cccc}
    0& F^{AB}_{\mathbf{k}} & G^{AA}_{\mathbf{k}} & 0 \\
    {[F^{AB}_{\mathbf{k}}]^*} & 0& 0 & {G}^{BB}_{\mathbf{k}} \\
    {[G^{AA}_{\mathbf{k}}]^*} & 0 &0 & {[F^{AB}_{\mathbf{k}}]^*} \\
    0 & {[G^{BB}_{\mathbf{k}}]^*} & {F}^{AB}_{\mathbf{k}} & 0
    \end{array}
    \right).
\eeq
was shown to maximally mix chiralities.

\begin{table}[]
\begin{tabular}{l|c c| c c}
\toprule
 Bond & $J_{xy} = J_{yx}$ & Splitting & $J_{xx} \neq J_{yy} $ & Splitting\\ 
\midrule
J1  & yes & no & no & -\\
J2  & yes & yes & no & - \\
J3  & yes & no & yes & yes \\
J4   & yes & no & yes & yes\\
J5  & yes & yes & no & -\\
J6 & yes & no & no & -\\
J7a/b & yes & yes & no &  -\\
\bottomrule
\end{tabular}
\caption{Allowed two-spin couplings that break U(1) symmetry}
\label{tab:bond_table}
\end{table}

Symmetric off-diagonal exchange $S_i^xS_j^y +  S_i^yS_j^x$ connecting identical sublattices is forbidden for the body-centered tetragonal structure but allowed for the rutile structure. Therefore it is a candidate for producing a net chirality. However, we find that it does not cause splitting of the magnon bands. This is because the bare couplings coupling A to A sublattices have a reversed sign compared to those coupling B to B. But, in the local quantization frame of the antiferromagnetic structure, the signs are the same and the coefficient is pure imaginary. There is therefore an effective translation symmetry connecting the two sublattices. 
\par

For the $J_{xx}\ \neq J_{yy}$ interaction for $3$rd  and $4$th neighbor bonds, there is splitting everywhere but the zone boundary and the $(HHL)$ and $(H\bar{H}L)$ planes. Once again this coupling does not lead to a finite chirality. 

In summary, we have examined all anisotropic exchange couplings on bonds out to $-$ and including $-$ seventh nearest neighbor. Of those that enter into the quadratic spin wave Hamiltonian, we have identified two classes that bring about a splitting of the two magnon bands. These are: $S_i^xS_j^y +  S_i^yS_j^x$ on $2$nd, $5$th and $7$th neighbor bonds and $S_i^xS_j^x -  S_i^yS_j^y$ on $3$rd and $4$th neighbor bonds. These do not lead to a net chiral term in the neutron cross section. In this way, we have shown that the sole two-spin coupling that generates a non-vanishing chiral term within linear spin wave theory is the imbalance between Heisenberg $J_7$ couplings.

\end{document}